\documentclass[namedreferences]{SolarPhysics}
\usepackage{epsfig}
\usepackage{courier}   

\newcommand{\etal}{{\it et al.}}
\newcommand{\adv}{{\it Adv. Space Res.}}

\newcommand{\aap}{{\it Astron. Astrophys.}}

\newcommand{\apj}{{\it Astrophys. J.}}
\newcommand{\apjl}{{\it Astrophys. J. Lett.}}

\newcommand{\grl}{{\it Geophys. Res. Lett.}}

\newcommand{\jastp}{{\it J. Atmos. Sol. Terr. Phys.}}
\newcommand{\jgr}{{\it J.~Geophys.~Res.}}

\newcommand{\pre}{ {\it Phys. Rev. E}}
\newcommand{\solphys}{{\it Sol. Phys.}}
\newcommand{\SpaceS}{{\it Space Sci. Rev.}}

\newcommand{\rmd}{ {\ \mathrm d} }

\newcommand{\dt}{~{\mathrm d}t}
\newcommand{\be}{\begin{equation}}
\newcommand{\ee}{\end{equation}}
\newcommand{\beq}{\begin{eqnarray}}
\newcommand{\eeq}{\end{eqnarray}}

\begin{document}
\begin{article}
\begin{opening}
\title{Progressive transformation of a flux rope to an ICME}
\subtitle{Comparative analysis using the direct and fitted expansion methods}
\author{S. \surname{Dasso}$^{1,2}$,
	M.S. \surname{Nakwacki}$^{1}$,
	P. \surname{D\'emoulin}$^{3}$,
	C.H. \surname{Mandrini}$^{1}$
}
\runningauthor{Dasso et al.}
\runningtitle{Transformation of a flux rope to an ICME}
\institute{
$^{1}$ Instituto de Astronom\'\i a y F\'\i sica del Espacio, CONICET-UBA,
CC. 67, Suc. 28, 1428 Buenos Aires, Argentina \email{sdasso@iafe.uba.ar}\\
$^{2}$ Departamento de F\'\i sica, Facultad de Ciencias Exactas y
Naturales, Universidad de Buenos Aires, 1428 Buenos Aires, Argentina\\
$^{3}$ Observatoire de Paris, LESIA, UMR 8109 (CNRS),
       F-92195 Meudon Principal Cedex, France\\}

\date{Received ; accepted }

\begin{abstract}
The solar wind conditions at one astronomical unit (AU)
can be strongly disturbed by the interplanetary
coronal mass ejections (ICMEs). A subset, called magnetic clouds (MCs),
is formed by twisted flux ropes that transport an important amount of
magnetic flux and helicity which is released in CMEs.
At 1~AU from the Sun, the magnetic structure of
MCs is generally modeled
neglecting their expansion during the spacecraft crossing.
However, in some cases, MCs present a significant expansion.
We present here an analysis of
the huge and significantly expanding MC observed
by the Wind spacecraft during 9 and 10 November, 2004.
This MC was embedded in an ICME.
After determining an approximated orientation for the
flux rope using the minimum variance method,
we precise the orientation of the cloud axis relating
its front and rear magnetic discontinuities using a direct method.
This method takes into account the
conservation of the azimuthal magnetic flux between
the in- and out-bound branches, and is valid for a finite impact
parameter (i.e., not necessarily a small distance between
the spacecraft trajectory and the cloud axis).
The MC is also studied using dynamic models
with isotropic expansion.
We have found $(6.2 \pm 1.5) \times 10^{20}$ Mx
for the axial flux, and
$(78 \pm 18) \times 10^{20}$ Mx for the azimuthal flux.
Moreover, using the direct method, we find that the ICME is formed by a
flux rope (MC) followed by an extended coherent magnetic region.
These observations are interpreted considering
the existence of a previous larger flux rope,
which partially reconnected with its environment in the front.
We estimate that the reconnection process started close to the Sun.
These findings imply that the ejected flux rope
is progressively peeled by reconnection and transformed
to the observed ICME (with a remnant flux rope in the front part).

\end{abstract}
\end{opening}

\section{Introduction}
\label{intro}

A magnetic configuration, previously in equilibrium
in the solar atmosphere, can reach a global
instability threshold
when the magnetic stress becomes too high.
In this case, the plasma is ejected
into the interplanetary (IP) medium and
is observed as a Coronal Mass Ejection (CME)
by the solar coronagraphs.
This magnetized mass, which can be expelled as fast as
few times $1000$~km/s, is recognized in the IP
space as an interplanetary CME, ICME,
see e.g., \inlinecite{Wimmer-Schweingruber06}.
During its travel from the Sun to 1 AU,
fast CMEs are slowed down due to drag forces between the
ICME and the solar wind environment
(see, e.g., \opencite{Vrsnak02}). Thus, at 1 AU
they can reach speeds as high as $\sim 1000$~km/s.

\subsection{Magnetic Clouds}
\label{Intro-MC}

Magnetic clouds (MCs) are a particular subset of ICMEs.
They are formed by twisted magnetic flux tubes that
carry a large amount of magnetic helicity from
the Sun to the IP medium. They also transport significant
amounts of magnetic flux, mass, and energy.
The principal characteristics of these magnetic structures are:
(i) an enhanced magnetic field,
(ii) a smooth rotation of the magnetic field vector through
a large angle ($\approx 180^{0}$), and
(iii) a low proton temperature \cite{klein82}.

The magnetic field in MCs is relatively well modeled
by the so-called Lundquist's model \cite{Lundquist50}, which
considers a static and axially-symmetric linear force-free
magnetic configuration (e.g., \opencite{Goldstein83}; \opencite{Burlaga88};
\opencite{Lepping90}; \opencite{Burlaga95}; \opencite{Lynch03}).
However, many other different models have been also used to describe the
magnetic structure of MCs.

Some models consider the MC as a rigid body during
the time it travels through the solar wind
and crosses the spacecraft.
\inlinecite{Farrugia99} considered a cylindrical
shape for the cloud cross-section and a non-linear force-free field;
while \inlinecite{Mulligan99}, \inlinecite{Hidalgo02}, and
\inlinecite{Cid02} considered a cylindrical cloud but a
non-force free field.
Non cylindrical static models have been
also applied to MCs (e.g., \opencite{Hu01}, \opencite{Vandas02}).
A comparison of global quantities (magnetic fluxes and helicity)
derived from different static models has been done
by \inlinecite{Dasso03} and \inlinecite{Dasso05}.
Different techniques
have been compared using synthetic data and analyzing the output of
numerical simulations of MCs \cite{Riley04}.

Some MCs present a significantly larger velocity
in their front than in their back, a characteristic of expansion.
Thus, some authors have used dynamical models to
describe these clouds during
the observation time; they have considered two cases:
only with a radial expansion
(see, e.g., \opencite{Farrugia93}, \opencite{Osherovich93b},
\opencite{Farrugia97}, \opencite{Nakwacki05}), and
with expansion in the radial and axial directions
(see, e.g., \opencite{Shimazu02}, \opencite{Berdichevsky03}).
Some dynamical models consider an expanding elliptical shape for
the MC (e.g., \opencite{Hidalgo03}).
The main aim of these models is to take into account the evolution
of the magnetic field as the spacecraft crosses the MC; then, to
correct the effect of mixing spatial-variation/time-evolution
in the observations to get a better determination
of the MC field (and related characteristics).

\subsection{Aims of this study}
\label{Intro-Aims}

We analyze the MC detected inside the ICME observed at L1 between
Nov.~9, 2004, at 20:25~UT and Nov.~11,
2004, at 18:45~UT \cite{Harra06}.
This is a very fast, left-handed and huge
MC with a size larger than
0.2~AU in the Earth-Sun direction. It has
a very intense magnetic field ($> 40$~nT), and expands strongly,
with a velocity of $\sim 850$~km/s in its front and
$\sim 600-700$~km/s in its back, depending on where the rear boundary is set.
This MC presents one of the largest (ever observed) velocity differences
between its front and its back \cite{Nakwacki07}.

We analyze this MC using a model-independent method,
called direct method \cite{Dasso06}. It takes into account the
magnetic flux conservation in closed structures, such as flux ropes.
This method gives us an estimation of the magnetic flux in the MC
directly from the data, and allows us to
improve the determination of the
orientation of the MC axis, as well as its boundaries.
Finding the boundaries for some MCs is an open issue
\cite{Russell05,Wimmer-Schweingruber06}. Indeed, several authors,
using different proxies, set them at different times with the consequent
differences in the
estimated MC axial orientation and in the
estimations of global magnetohydrodynamic quantities.

In Section~\ref{Section_MC}, we describe the
magnetic and plasma properties of the MC.  Then, in
Section~\ref{Section_direct_method}, we present the direct method and
the results obtained.  In
Section~\ref{Sect_Models}, we fit the two dynamical models to velocity
and magnetic field observations.  From the fitted models
we compute the magnetic fluxes, and compare them with the values
obtained from the static Lundquist's model and from the direct method.
Finally, in Section~\ref{Conclusion}, we give a summary and our
conclusions.\\

\section{The studied cloud}
\label{Section_MC}
 In this section we analyze {\it in situ} magnetic and plasma
observations of the MC observed on Nov. 9-10, 2004, which is located inside an
ICME.  We define a local frame of coordinates, attached to the MC,
and we analyze the data in this frame.

\subsection{One or two clouds?}
\label{MC_1or2}

\begin{figure}
 \centerline{\includegraphics[width=1.0\linewidth]{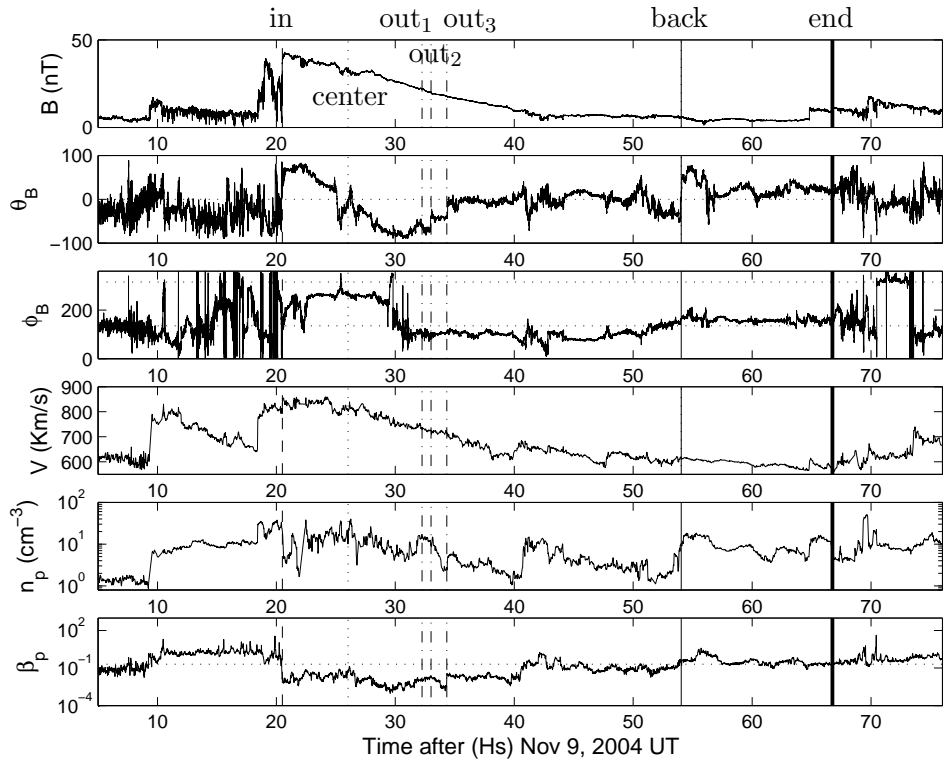}}
 \vspace{-0.81\linewidth}
 \centerline{\hspace{0.275\linewidth} in
             \hspace{0.07\linewidth} out$_1$
             \hspace{-0.01\linewidth} out$_3$
             \hspace{0.14\linewidth} back
             \hspace{0.085\linewidth} end  \hfill
            }
 \centerline{\vspace{0.01\linewidth} \hspace{-0.09\linewidth} out$_2$
            }
 \centerline{\hspace{0.32\linewidth} center \hfill}
  \vspace{0.72\linewidth}
\caption{
Wind observations for the magnetic cloud observed inside the
ICME of 9-11 Nov. 2004 (time cadence of 100 seconds).  From upper
to lower panels: absolute value of the magnetic field ($B=|\vec{B}|$),
latitude ($\theta_B$) and longitude ($\phi_B$) angles of
$\vec{B}$ in GSE, the bulk velocity ($V$), the proton density
($n_p$), and
the proton plasma beta ($\beta_p$), all of them as a function of time.
The vertical dashed
line ``in'' corresponds to the start of the field coherent rotation
(beginning of the MC, Nov.~9 at 20:30~UT), the vertical dotted line
to the cloud center (Nov.~10 at 02:02~UT,
see Section~\ref{direct_method_structure}), the vertical
dash-dotted lines ``out$_1$'' and ``out$_3$''
to the extremes of the range of
possible endings for the rotation of $\vec{B}$
(end of the cloud, Nov. 10 at 08:15~UT and 10:20~UT), and
the dash-dotted line ``out$_2$'' to a strong
discontinuity (in between this range,
on Nov.~10, at 09:00~UT).
The thin solid vertical line ``back'' marks a strong magnetic
discontinuity (Nov.~11, at 06:02~UT) and
the thick vertical solid line ``end'' the end of the ICME
(Nov.~11, at 18:45~UT).  Horizontal dotted lines in $\theta_B$ and
$\phi_B$ panels indicate the orientation of Parker's spiral, while the one in
$\beta_p$ panel marks the mean value for a set of MCs studied by
Lepping {\it et al.} (2003).
}
\label{fig_BV_gse}
\end{figure}

  An ICME was observed from 20:25~UT on 9 Nov. to $\approx$~18:45~UT on
11 Nov. The ICME is preceded by a typical piled up solar wind material
(panel $n_p$ in Figure~\ref{fig_BV_gse}).  This corresponds to plasma and
magnetic field pushed from behind by the ICME, forming the turbulent
pre-ICME sheath (notice the high level of fluctuations in $\theta_B$).
A forward shock is located in front of the sheath (at 9:15~UT of
Nov.~9).  This event presents also a non-typical second shock inside
the sheath that precedes the ICME (at 18:20~UT of Nov.~9).  For a deeper
description of shocks and their association with solar sources
see \inlinecite{Harra06}, in particular their Table~I.

The end of the ICME is marked by a thick solid line (Nov.~11,
18:45~UT, label ``end'' in Figure~\ref{fig_BV_gse}) as defined
by \inlinecite{Harra06}.  After Nov.~11 at
18:45~UT, the magnetic field is consistent with Parker's spiral
(changes of magnetic sectors are observed in $\phi_B$).  The higher
level of magnetic fluctuations, typical of the fast solar wind, confirms this
interpretation.

Two MCs were initially reported inside this ICME:
the first one starting on Nov.~09, at 20:54~UT, and finishing on
Nov.~10, at 03:24~UT, and the second one starting on Nov.~10, at
03:36~UT, and finishing on Nov.~10 at 11:06~UT
(\texttt{http://\-lepmfi.gsfc.nasa.gov/\-mfi/\-mag\_cloud\_S1\-.html}).
The rotation of the field is indeed larger than usual (close to one
turn, Figure~\ref{fig_BV_gse}). Moreover, the magnetic field observed
has indeed a non-classical structure: it is very strong in the front,
progressively decreases in the back and has an extended weak tail.
Still, a coherent and continuous variation of both the field strength
and the plasma velocity is
observed, without clear evidence of two independent magnetic structures.

Other studies concluded that only one extended MC was present.
In \inlinecite{Harra06} and \inlinecite{Longcope07} the
boundaries of the magnetic cloud were selected as starting on Nov.~9,
at 20:30~UT, and finishing on Nov.~10, at 10:00~UT.  Similar
boundaries (starting on Nov.~9, at 20:40~UT, and finishing on Nov.~10,
at 10:20~UT) were chosen by \inlinecite{Qiu07}. Both the direct method
and data modeling confirm the presence
of only one flux rope (Sections~\ref{Section_direct_method}
and~\ref{Sect_Models}).

\subsection{Summary of the cloud observations}
\label{MC_obsGSE}

We analyze the {\it in situ}
measurements of the magnetic field components obtained by the
Magnetic Field Instrument (MFI, \opencite{Lepping95b}) and the
plasma quantities obtained by the Solar
Wind Experiment (SWE, \opencite{Ogilvie95}), both aboard Wind.
There is a small data gap in SWE from Nov.~9, 21:58~UT to
22:27~UT (see Figure~\ref{fig_BV_gse}).

The magnetic field observations are in GSE (Geocentric Solar Ecliptic)
coordinates. In this right-handed system of coordinates, $\hat{x}_{GSE}$
corresponds to the Earth-Sun direction, $\hat{z}_{GSE}$ points to the
North (perpendicular to the ecliptic plane) and $\hat{y}_{GSE}$ is in
the ecliptic plane and points to the dusk when an observer is near
Earth (thus, opposing to the planetary motion).

The front boundary of the MC (Nov.~9, at 20:30~UT) is well defined
(vertical dashed line in Figure~\ref{fig_BV_gse}, label ``in'').
The magnetic field
presents a North-West-South rotation with time (see $\theta_B$ and $\phi_B$
panels in Figure~\ref{fig_BV_gse}); thus, the MC is formed by a left-handed
flux rope with its axis almost on the ecliptic and pointing to
the west ($y_{GSE}<0$ or $\phi_B \sim 270^\circ$).

A characteristic of this MC is its very strong expansion (see panel
$V$ in Figure~\ref{fig_BV_gse}).  The observed plasma
velocity, $V$, goes from $\sim 850$~km/s at the beginning to $\sim
700$~km/s close to the cloud end, a difference of $150$~km/s in the
observed time range ($\sim 15$ hours).  This implies an expansion
of $\sim 10$~km/s per hour.

The MC rear boundary is uncertain, the vertical dash-doted lines in
Figure~\ref{fig_BV_gse} indicate a possible range (from Nov.~10 at
8:15~UT to 10:20~UT, labels ``out$_1$'' and ``out$_3$'', respectively).
However, the decrease of $|\vec{B}|$ and the strong expansion are
still present at later times indicating a backward extension of the
MC. The mean value of $\beta_p$
(ratio of the proton pressure to the magnetic pressure) is 0.12
for a sample of MCs studied by
\inlinecite{Lepping03} (horizontal dotted line in panel
$\beta_p$ of Figure~\ref{fig_BV_gse}).  Then, another indication
that the MC is more extended in the back is the presence of
$\beta_p<0.12$ after 10:20~UT.  This region extends up to a
strong discontinuity in $\theta_B$ and density (on Nov.~11 at 06:02~UT,
label ``back'' in Figure~\ref{fig_BV_gse}). We call the region
between ``out$_3$'' and ``back'' simply the back region of the MC.
Finally, there is a region with weak, but coherent, field between the ``back''
and ``end'' boundaries. The physical origin of these regions is
analyzed in Section~\ref{direct_method_ICME_structure}.

\subsection{Orientation and extension of the cloud}
\label{MC_orientation}

 To facilitate the understanding of the MC properties, we
define a system of
coordinates linked to the cloud in which $\hat{z}_{\rm cloud}$ is
along the cloud axis (with $B_{z, \rm cloud}>0$).
  We define the latitude angle ($\theta$) between the ecliptic plane and
the cloud axis, as well as the longitude angle ($\varphi$) between the
projection of the axis on the ecliptic plane and the Earth-Sun
direction ($\hat{x}_{GSE}$) measured counterclockwise.
  Then, when $\theta$=90$^\circ$ ($\theta$=-90$^\circ$)
the cloud axis is parallel (antiparallel)
to $\hat{z}_{GSE}$ and it points to the
ecliptic North (South).  When $\theta$=0$^\circ$ the cloud axis is on
the ecliptic plane, $\varphi$=0$^\circ$ being the case of the axial
field pointing toward the Sun, and $\varphi$=90$^\circ$
($\varphi$=270$^\circ$) when it points to the terrestrial dusk (dawn).

  Since the speed of an MC is nearly in the Sun-Earth direction and
it is much larger than the spacecraft speed (which can be supposed to
be at rest during the cloud observing time), we assume a rectilinear
spacecraft trajectory in the cloud frame.  The trajectory defines a
direction $\hat{d}$ (pointing toward the Sun);
then, we define $\hat{y}_{\rm cloud}$ in the
direction $\hat{z}_{\rm cloud} \times \hat{d}$ and $\hat{x}_{\rm
cloud}$ completes the right-handed orthonormal base ($\hat{x}_{\rm
cloud},\hat{y}_{\rm cloud},\hat{z}_{\rm cloud}$).
We also define the impact parameter, $p$, as the minimum
distance from the spacecraft to the cloud axis.
Then, we construct a rotation matrix from the GSE system to the cloud
system and obtain the components of the observed magnetic field in
the cloud coordinates: {$B_{x, \rm cloud}$, $B_{y, \rm cloud}$, $B_{z,
\rm cloud}$}.

The local system of coordinates is especially useful when $p$ is small
compared to the MC radius ($R$).  In particular, for $p=0$ and an MC
described by a cylindrical magnetic configuration $\vec{B}(r) = B_z(r)
\hat{z} + B_\phi(r) \hat{\phi}$, we have $\hat{x}_{\rm cloud} =
\hat{r}$ and $\hat{y}_{\rm cloud} = \hat{\phi}$ when the spacecraft
leaves the cloud.
In this particular case, the magnetic field data will
show: $B_{x, \rm cloud}=0$, a large and coherent variation of $B_{y,
\rm cloud}$ (with a change of sign), and an intermediate and coherent
variation of $B_{z, \rm cloud}$, from low values at one cloud edge,
taking the largest value at its axis and returning to low values at
the other edge.

The minimum variance (MV) method \cite{Sonnerup67} has been used to
estimate the orientation of MCs (see e.g., \opencite{Bothmer98},
\opencite{Lepping90}, \opencite{Farrugia99}, \opencite{Dasso03},
\opencite{Gulisano05}).  It gives a good estimation
if $p$ is small compared to $R$ and if the
in-/out-bound magnetic fields are not significantly asymmetric.

However, when the cloud presents a strong expansion, as in the event
studied here, the directions derived by the MV
method will mix two different effects in the variance of $\vec{B}$:
(1) the effect of the coherent rotation of $\vec{B}$ (which provides
the cloud orientation), and (2) the effect of the cloud
'aging' (the decrease of the field strength with time
due to magnetic flux conservation combined with cloud expansion).
This latter effect is not associated with the cloud orientation;
thus, we apply the MV technique to $\vec{B}/B$ to
decrease the cloud 'aging' consequences.

 We start the analysis taking the ``in'' and
the ``out$_3$'' boundaries, since the strongest
magnetic discontinuity
is located at ``out$_3$''.
With these boundaries, we find the typical shape of the
components of $\vec{B}$ in the cloud frame, as discussed above
(see Section~\ref{direct_method_structure} for
further justifications).
The MV method applied to the normalized field
gives $\theta$=-23$^\circ$ and
$\varphi$=274$^\circ$.
With the same procedure and an end boundary on
Nov. 10, 10:00UT, \inlinecite{Harra06} found $\theta$=-20$^\circ$ and
$\varphi$=276$^\circ$.
Changing the cloud end between ``out$_1$'' and ``out$_3$'',
the ranges for $\theta$ and $\varphi$ are:
$\theta \sim$[-25,0]$^\circ$ and $\varphi
\sim$[260,280]$^\circ$.

\begin{figure}
 \centerline{\includegraphics[width=1.0\linewidth]{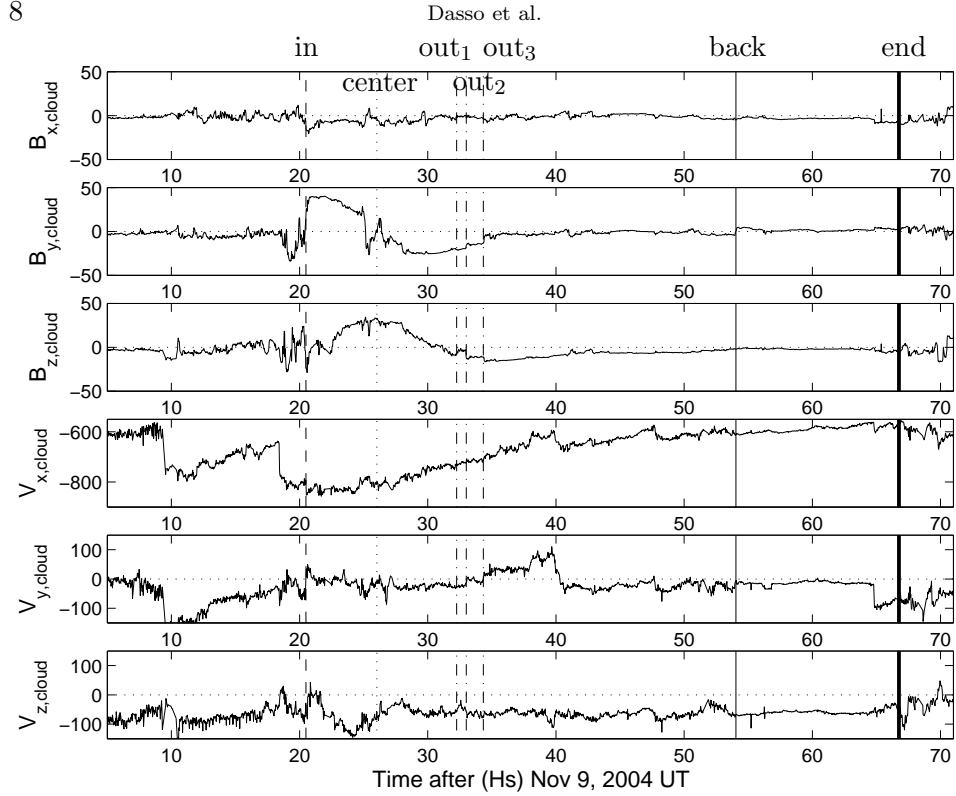}}
 \vspace{-0.815\linewidth}
 \centerline{\hspace{0.29\linewidth} in
             \hspace{0.085\linewidth} out$_1$
             \hspace{-0.01\linewidth} out$_3$
             \hspace{0.16\linewidth} back
             \hspace{0.1\linewidth} end  \hfill
            }
 \centerline{\vspace{0.01\linewidth} \hspace{-0.02\linewidth} out$_2$
            }
 \vspace{-0.045\linewidth}
 \centerline{\hspace{0.34\linewidth} center \hfill}
  \vspace{0.75\linewidth}
\caption{From upper to lower panels, observed magnetic and velocity
field components in the cloud frame. In this frame $\hat{z}_{\rm cloud}$ is
along the cloud axis ($\theta=-10^\circ$ and $\varphi=275^\circ$
in GSE coordinates), $\hat{y}_{\rm cloud}$ is orthogonal to both
the MC axis and the spacecraft trajectory, and $\hat{x}_{\rm cloud}$
completes the right-handed orthogonal base. Vertical lines
correspond to the same times as in Figure~\ref{fig_BV_gse}. The region
between ``in'' and  ``center'' is called the in-bound, and
between ``center'' and ``out$_1$'' or ``out$_2$'' or ``out$_3$''
the out-bound. From ``out$_1$'' or ``out$_2$'' or ``out$_3$''
up to ``back'' we have the back region,
while after that and up to ``end'' completes the ICME extension.}
\label{fig_BV_cloud}
\end{figure}

\subsection{The data in the cloud frame}
\label{MCframe_data}

Figure~\ref{fig_BV_cloud} shows the components of the magnetic and
velocity field in the cloud frame for an orientation of the cloud axis
such that: $\theta=-10^\circ$ and $\varphi=275^\circ$ (this
orientation is justified in Section~\ref{direct_method_orientation}).
The magnetic field components show the typical large scale shape of MCs when
the impact parameter is small compared with the cloud radius.
However, an inner and non-typical sub-structure is present at its
center (mainly observed in $B_{y,cloud}$, where it is antisymmetric).

A strong expansion is observed in the velocity components,
mainly along $\hat{x}_{\rm cloud}$.  This is expected from the MC
orientation and the data in GSE since, as $\varphi \sim 270^\circ$, the
spacecraft cannot observe different parcels of fluid along a large
range of $z_{cloud}$ values.
However, a weak signature of expansion along the cloud axis
can be observed in $V_{z,cloud}$
(Figure~\ref{fig_BV_cloud}).  For an MC
with $\theta \sim 0$ and $\varphi > 270^\circ$, an axial expansion is
characterized by $V_{z,cloud}<0$ in the front, changing to
$V_{z,cloud}>0$ in the back.  This is just what is observed in our case
when the mean value of $V_{z,cloud}$ is removed within the cloud.
This gives us the clue that $\varphi$ is larger than 270$^\circ$.

When the spacecraft crosses a cylindrical MC
(or an elliptical one with one of the main axis parallel
to the Sun-Earth direction) and $p=0$,
$B_{x,cloud} \approx 0$.
The first panel in Figure~\ref{fig_BV_cloud} shows that
the observed $B_{x,cloud}$ has a slightly negative mean value.
The sign of $B_{x,cloud}$, together with the evolution of
$B_{y,cloud}$, implies that
the flux rope axis is above the ecliptic plane.

\section{Results with the Direct Method}
\label{Section_direct_method}

\subsection{The Direct Method}
\label{the_direct_method}

In this section we summarize and extend the direct method presented by
\inlinecite{Dasso06}. This method lets us find the
rear boundary of a flux rope for a given axis orientation,
or the reverse, the MC orientation for a given position of the rear boundary.
The front boundary of a flux rope is usually well defined
by a discontinuity
of the magnetic field (changing abruptly from a fluctuating field in the
MC sheath to a strong and coherent field within the MC).
The corresponding
current sheet is expected to be present all around the flux rope
as the limit between magnetic regions with different magnetic
connectivities; so, with different magnetic stresses.
A corresponding magnetic discontinuity, labeled ``out'',
is then expected at the rear of the MC.
The flux rope is present in between these two discontinuities and the
same amount of azimuthal flux is traversed
twice by the spacecraft.

Let us consider a flux rope at a given time.
The conservation of the magnetic flux
($\vec{\nabla} \cdot \vec{B}=0$)
across a plane formed by the spacecraft trajectory
and $\hat{z}_{cloud}$ ($y_{cloud}$ constant)
gives \cite{Dasso06}:
    \begin{equation} \label{Byflux}
    \int_{\rm flux~rope}  B_{y, \rm cloud} \rmd x \rmd z = 0 \,,
    \end{equation}
with $x, z$ being
the spatial coordinates in the
$\hat{x}_{cloud}$ and $\hat{z}_{cloud}$ directions,
respectively.

The observations provide only $B_{y, \rm cloud}$ as a
function of time along the trajectory. So, we need two
hypothesis: an invariance of $B_{y, \rm cloud}$
along the flux rope axis and the conservation of the magnetic flux with time.
The first hypothesis is justified by a low ratio of
the MC radius over the expected curvature radius of the axis and
the balance of magnetic torques, which is expected to homogenize the field
along the axial direction \cite{Dasso06}.
The second hypothesis is valid as far as the amount of magnetic flux
reconnected during the spacecraft crossing is low.
Indeed, we have found that magnetic flux is reconnected in the front of the MC
(see Section~\ref{direct_method_ICME_structure}). An estimation of the amount
of flux reconnected during the spacecraft crossing gives $\sim$ 5\% of
the initial azimuthal flux
(see Section~\ref{direct_method_reconnection}); so, unless the reconnection
rate is much higher during the observing time than it was during the travel
from the Sun, the amount of reconnected flux is small
during the crossing. We neglect below such reconnected flux.

 The elementary flux crossed during ${\mathrm d}t$ is
$B_{y,cloud}(t) \, L(t) \, V_{x,cloud}(t) \, \rmd t$, where $L(t)$ is
the axial length of the portion of the flux rope which had a length
$L_{\rm in}=L(t_{\rm in})$ when the spacecraft entered the MC. Then,
Equation~(\ref{Byflux}) becomes:
    \begin{equation} \label{Byflux2}
    \int_{\rm flux~rope}  B_{y,cloud}(t) \ L(t) \ V_{x,cloud}(t) \rmd t = 0 \,.
    \end{equation}

If the axis orientation and the position of one
boundary of an MC are known, the above flux balance
property can be used
to find the MC center and the other boundary as follows. We define
the accumulative flux per unit length:
    \begin{equation} \label{Byaccumul}
    \frac{F_y(x)}{L_{\rm in}} = \int_{t_{\rm in}}^{t(x)} B_{y,cloud}(t') \
                           \frac{L(t')}{L_{\rm in}} \ V_{x,cloud}(t')
                           \  \rmd t' \,,
    \end{equation}
where $t_{\rm in}$ is the time of the MC front boundary (located at
$x=X_{\rm in}$) and $x=\int_{t_{\rm in}}^{t} V_{x,cloud}(t') \rmd t'$.
The position where $F_y(x)/L_{\rm in}$ has its
absolute extreme gives an estimation of the position where the spacecraft
reaches the closest distance to the MC axis.  This indicates the
$x$ position of the MC center, being
this estimation more precise as the impact parameter is lower.
Then, when $F_y(x) /L_{in}$ goes back to zero at $x=X_{\rm out}$,
we have the other boundary.  The region from $x=X_{\rm in}$ to
$x=X_{\rm out}$ defines the MC flux rope.

  Since the ratio of the MC radius to the Sun distance is small (typically
$\approx 0.1$ or lower), $L(t')/L_{\rm in} \approx 1$. Below we derive
a correction to this estimation. If the MC axis does not change
drastically its shape,
and if it does not disconnect from the Sun
during the crossing time, its length is evolving
proportionally to its distance to the Sun, $D(t)$, so
$L(t)/L_{\rm in}=D(t)/D_{\rm in}$
(where $D_{\rm in} \approx 1$~AU for the present observations).
Moreover, the cloud global velocity,
which is the velocity of its center $V_{c}$, is not expected to
change significantly during the crossing time; then, we have:
    \begin{equation} \label{Length_evol}
    \frac{L(t)}{L_{\rm in}} \approx \frac{D(t)}{D_{\rm in}}
     = 1 + (t-t_{\rm in}) \frac{V_{c}}{D_{\rm in}}  \,.
    \end{equation}
All terms in the right hand side of Equation~(\ref{Byaccumul})
can be derived from observations.

\begin{figure}
 \centerline{\includegraphics[width=0.5\linewidth]{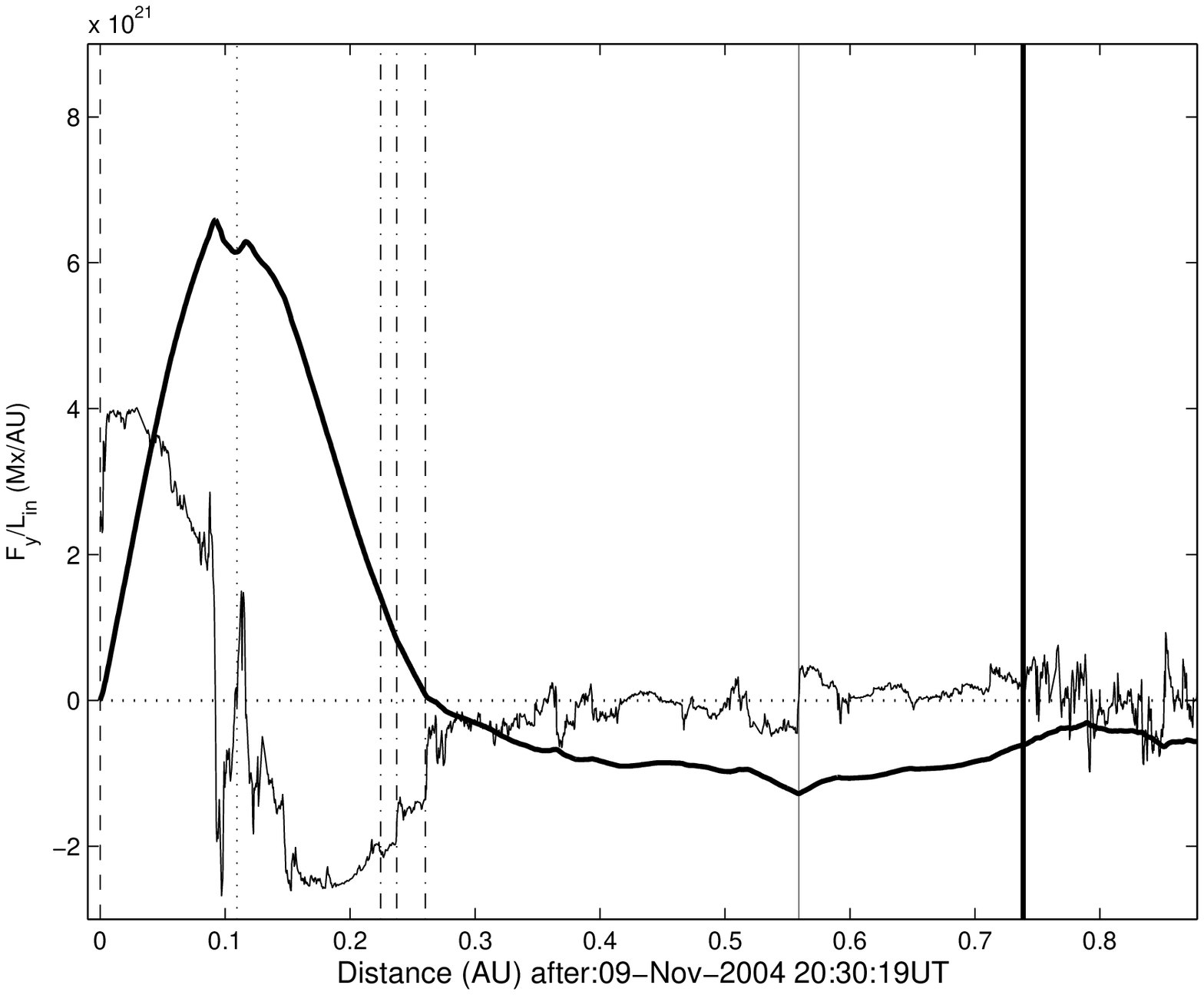}
             \includegraphics[width=0.5\linewidth]{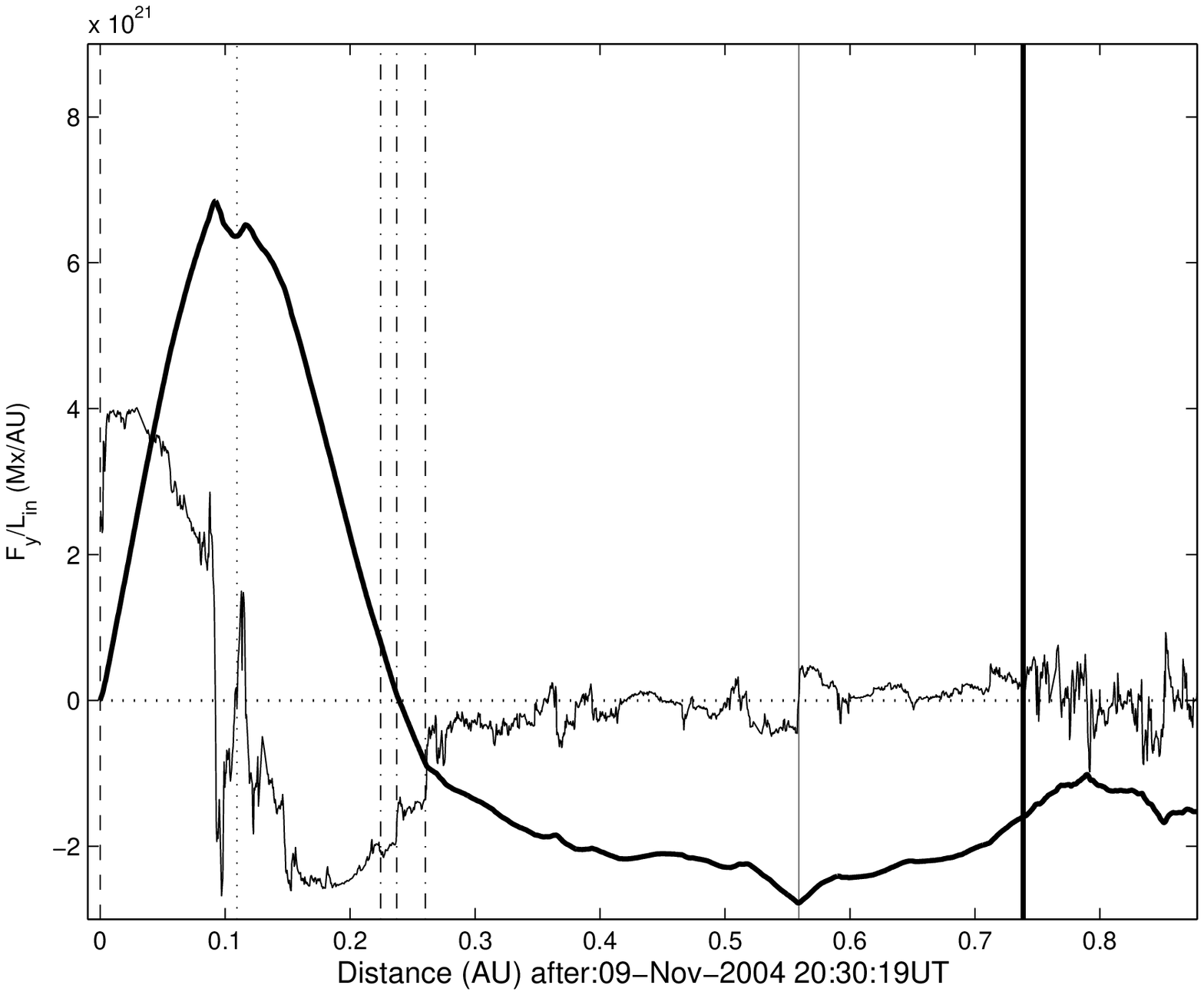}
            }
\caption{
Component of $\vec{B}$ perpendicular to both the trajectory
of the spacecraft and to the cloud axis ($B_{y,cloud}$, thin line curve)
and accumulated magnetic flux of this component per unit length
($F_y/L_{\rm in}$, thick line curve). The dotted vertical line marks the
cloud center (Nov.~10, 02:02 UT).
Left and right panels correspond to computation
of $F_y/L_{\rm in}$ without ($L(t)=L_{\rm in}$) and with axial expansion
(Equation~\ref{Length_evol}), respectively.
In both panels the cloud orientation was taken
as $\theta=-10^\circ$ and $\varphi=275^\circ$.
The vertical lines indicate the same positions as in
Figures~\ref{fig_BV_gse} and~~\ref{fig_BV_cloud}.
}
\label{fig_Fy}
\end{figure}

\subsection{Refined orientation of the cloud}
      \label{direct_method_orientation}

 The direct method was previously applied to the Oct. 18-20, 1995, MC
\cite{Dasso06}. The orientation of its axis was well determined.
The strong frontal discontinuity in the magnetic field
was naturally related to another
strong backward discontinuity, the flux balance given by
Equation~(\ref{Byflux2}) was satisfied, and $L(t)$ was constant.

  The orientation of the axis of the
MC studied here is not so well determined; then, we explore
different orientations to find which angles ($\theta$ and $\varphi$)
give a cancellation of $F_y$ at the strongest magnetic discontinuity
observed at the MC rear (labeled ``out$_{3}$''
in Figures~\ref{fig_BV_gse},~\ref{fig_BV_cloud}), at the inner extreme
labeled ``out$_{1}$'', and at the intermediate time ``out$_{2}$''
(between ``out$_{1}$'' and ``out$_{3}$'').
Due to the orientation of the MC (i.e.,
its axis almost lying on the ecliptic and perpendicular
to the Sun-Earth direction) $F_y$ is mostly affected by the value of
$\theta$.
We find $\theta =-10^\circ\pm 10^\circ$ from variations of the
end boundaries in the full range between ``out$_{1}$'' and ``out$_{3}$'',
and using the two extreme possibilities on the axial expansion
(no axial expansion, $L(t)=L_{in}$, and axial expansion proportional
to the distance to the Sun, Equation~\ref{Length_evol}).
The value of $\varphi$ is constrained
by imposing that $B_{x,cloud}(t)$ should have
a small variation with time (as expected in flux rope models);
so, with no contribution of
azimuthal or axial field components.
This gives $\varphi = 275^\circ\pm 10^\circ$.

  Left panel of Figure~\ref{fig_Fy} shows $F_y$
for $L(t)=L_{in}$ (no axial expansion)
and an orientation such that $F_y$ is canceled at the discontinuity
``out$_{3}$'', which gives $\theta =-10^\circ$ and $\varphi = 275^\circ$.
 The rear boundary of the MC is in fact ambiguous from the data,
since there is also a
strong discontinuity between
``out$_{1}$'' and ``out$_{3}$'' (``out$_{2}$'', at Nov 10, at 09:00UT),
see Figures~\ref{fig_BV_gse} and ~\ref{fig_BV_cloud}.
Fixing the previous orientation, but using the axial expansion given in
Equation~(\ref{Length_evol})
(see panel in Figure~\ref{fig_Fy}), we find that the cancellation
of $F_y$ is now at ``out$_{2}$''.
With this boundary, we find a much less extended region with a reversal
of $B_{z, \rm cloud}$ at the back of the MC. So, this
boundary gives an MC field closer to the one inferred from a classical
MC model. In this case, the results of the direct method
agree with those of the fitted models (including expansion,
Section~\ref{Sect_Models}). These are evidences that we have identified
the right end boundary and orientation of the flux rope.

\subsection{Structure of the cloud}
      \label{direct_method_structure}

\begin{figure}
 \centerline{\includegraphics[width=0.5\linewidth]{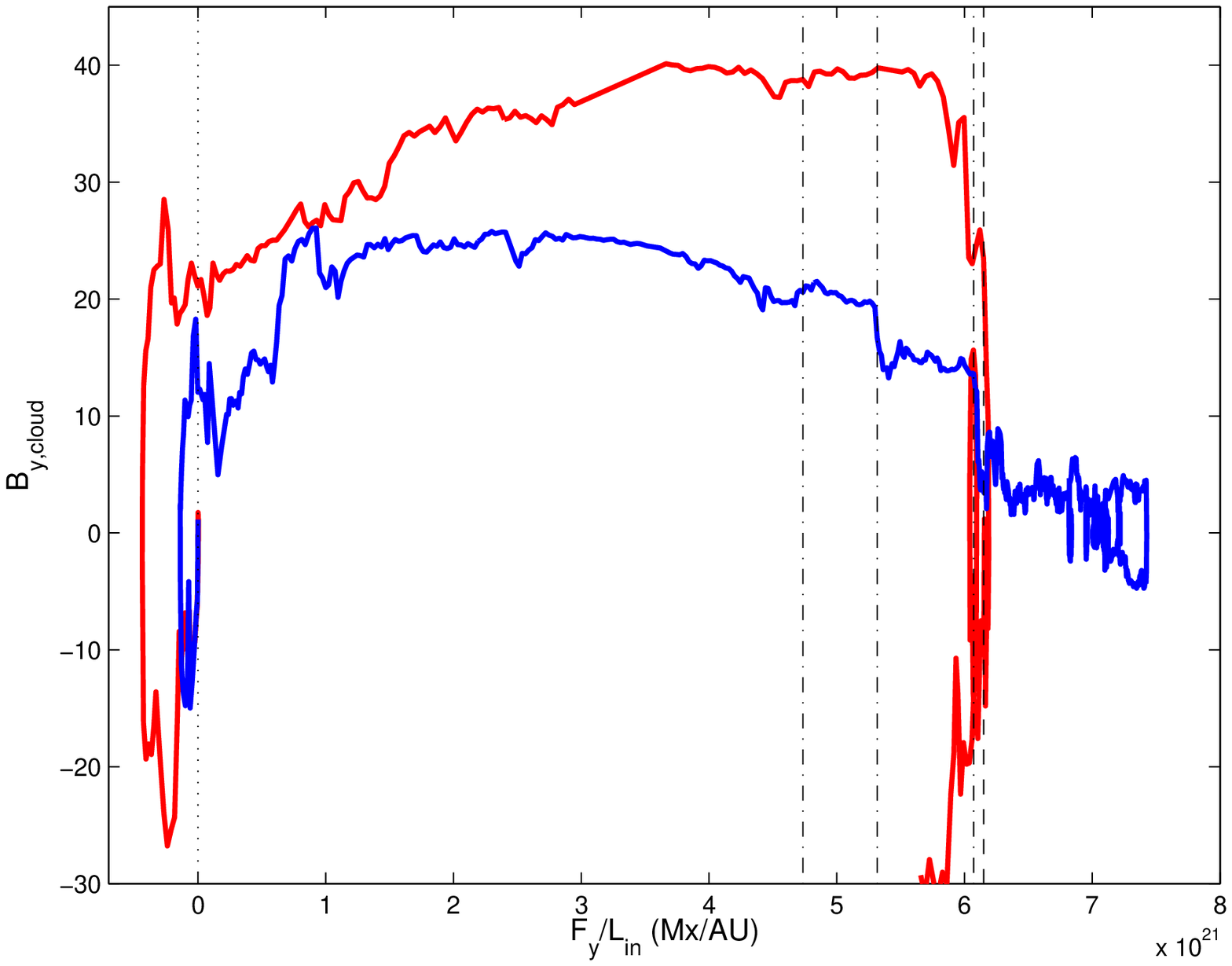}
             \includegraphics[width=0.5\linewidth]{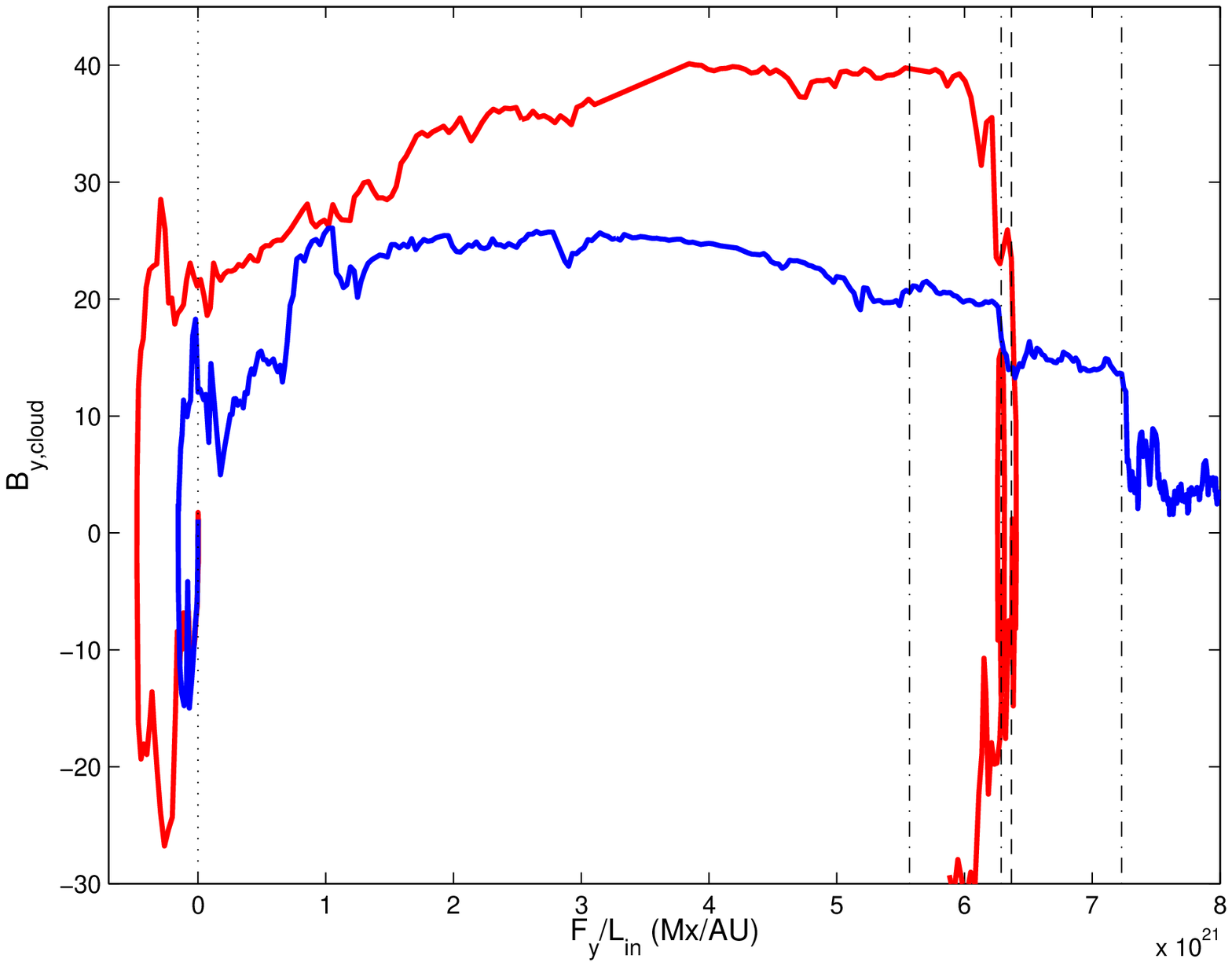}
            }
\caption{In-bound (red) and out-bound (blue) $y_{cloud}$ component of $\vec{B}$
in function of
the accumulated magnetic flux of this component per unit length
($F_y/L_{\rm in}$). $F_y/L_{\rm in}$ is computed as
in Figure~\ref{fig_Fy}.
The difference in the amplitude of
$B_{y,cloud}$ between the in-/out-bound regions
is due to the MC aging (not corrected here).}
\label{fig_By_vs_Fy}
\end{figure}

  The field component $B_{y, \rm cloud}$ vanishes at three locations
near the cloud center (close to the
vertical dotted line at the abscissa $\sim$ 0.1 AU in
Figure~\ref{fig_Fy}).  Taking into account the expected antisymmetry
of the in- and out-bound regions,
we set the cloud center at Nov.~10 at 02:02~UT
(dotted line in Figure~\ref{fig_Fy}).
Then, the accumulated flux $F_y/L_{\rm in}$ (Equation~\ref{Byaccumul})
gives a unique relationship between
the in- and out-bound data, since it labels each flux surface.  This
relationship is better shown using $F_y/L_{\rm in}$ in the abscissa and
reversing the sign of $B_{y,cloud}$ inside the out-bound branch
(Figure~\ref{fig_By_vs_Fy}).  Then, peaks and valleys of $B_{y, \rm cloud}$
in the in-/out-bound branches can be related.  As expected, this
association is stronger near the MC center where the regions that are crossed
are closer and also more isolated from the
interaction with the solar wind environment \cite{Dasso05b}. Except for the
strong discontinuity at the MC borders, the association between
structures is not clear outside the core (where
$B_{y, \rm cloud}$ has no characteristic variations
that can be recognized in both the in- and out-bound branches).

\begin{figure}
 \centerline{\includegraphics[width=1.\linewidth]{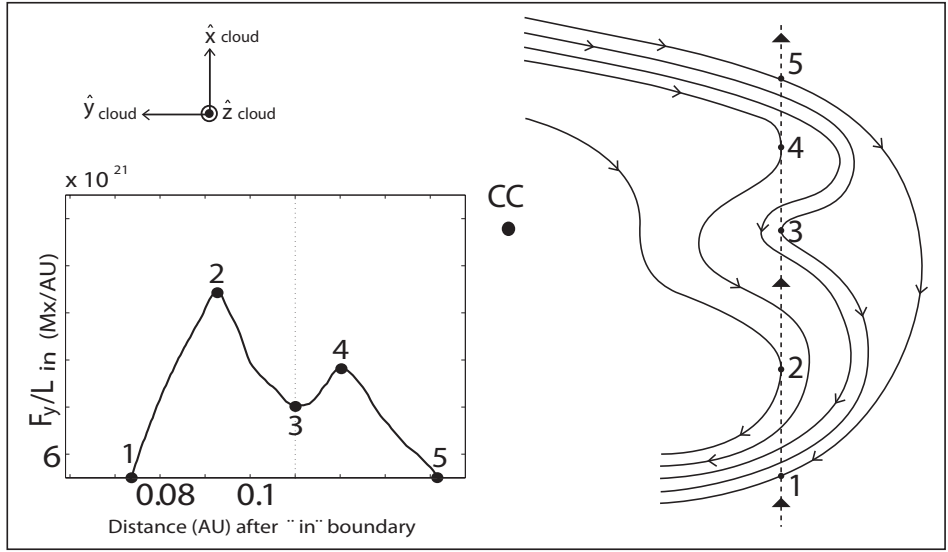}}
\caption{
  Scheme of the magnetic field lines
near the MC central region (to the right) and
the corresponding $F_y/L_{in}$ evolution along the spacecraft
trajectory (low left figure, a zoom of Figure~\ref{fig_Fy}).
The magnetic structure of the
flux rope is undulated, probably due to its fast evolution
and interaction with the surrounding medium. The
undulations are amplified in this scheme for clarity,
but in the observed
MC they are significant enough to produce reversals of $B_{y, \rm cloud}$.
Dot at CC indicates the location of the cloud center.
Numbers 1, 2, 3, 4 and 5 are reference points of the observed
field lines.}
\label{fig_Center}
\end{figure}

 There is a clearly distinguishable sub-structure in the cloud center
seen as reversed peaks in $B_{y, \rm cloud}$ and as a
valley in $F_y/L_{\rm in}$ (Figure~\ref{fig_Fy}). It has a small extension,
$\sim \pm 10^{-2}$~AU, and is globally anti-symmetric.
A less evident sub-structure was also present in the previously
analyzed MC (18 Oct. 1995, \opencite{Dasso06}).

  If $p=0$, the central sub-structure would
imply the presence of a small twisted flux tube with opposite magnetic
helicity in the center of the flux rope. The formation of such a structure
is not possible in the corona. After analyzing several
possibilities, the simplest interpretation is the following.
Close to the minimum approach, the spacecraft trajectory is nearly tangent
to the magnetic flux surfaces of the flux rope ($B_{y, \rm cloud} \approx 0$).
Any warping of the flux surfaces gives a clear signal in the
$B_{y, \rm cloud}$ component. For geometrical reasons, such warping
is more difficult to detect outside the center (where $B_{y, \rm cloud}$
is important). It is noteworthy that such structure will be evident
only if the magnetic data are rotated to the correct MC frame
to have no mixing with the strong $B_{z, \rm cloud}$ component.

  For the 18 Oct. MC, the warping was moderate and $B_{y, \rm cloud}$
kept its sign inside both the in- and the out-bound regions.
For the Nov.~9-10 MC, the warping
is more marked as sketched in Figure~\ref{fig_Center}.
This figure shows (to the right) a scheme of the spacecraft
trajectory (vertical dashed line)
across the MC core field lines (solid warped lines).
The core crossing starts at field line '1' and ends at '5'.
The large dot called 'CC' marks the cloud center
position. It is located towards positive values of $y_{cloud}$
since the cloud axis is above the ecliptic plane (see last paragraph of
Section~\ref{MCframe_data}).
Considering the cloud as a flux rope,
which is compatible with the observations,
the same accumulated flux $F_y/L_{in}$ implies
that the in- and out-bound field lines are connected
(i.e., they are in fact the same field line observed
twice). Thus, for example '1' is connected to '5'.
The flux $F_y/L_{in}$ has a local maximum at '2' and '4' and a local minimum
at '3' when the spacecraft trajectory is tangent to the field lines.

\subsection{Magnetic fluxes}
      \label{direct_method_flux}

  The total azimuthal flux $F_{\phi}$ is estimated taking $F_y(x_{center})$,
assuming that $p=0$. The largest source of uncertainty is the
length of the flux rope, $L_{\rm in}$, and in a more general way
the assumed invariance
by translation. Since this cloud presents signatures of being detached
at one of its legs \cite{Harra06}, we assume an initial length
$L_{\rm in}=1.5$~AU, so intermediate between values used in previous
papers.
The uncertainty in the MC boundary of the out-bound branch has a lower
effect since the difference of $F_{\phi}$ found with the boundary
``out$_{1}$'' and ``out$_{3}$'' is only about 10\% (see Table~\ref{tab_flux}).

The axial flux $F_z$, across a surface perpendicular to the
cloud axis, can be estimated directly from the
observations assuming a circular MC cross section,
$p=0$ and neglecting the expansion to compute $r=x(t)-x_{\rm center}$.
Here, we neglect the axial flux in the core since it is a correction
of the order of $(p/R)^2$, see \inlinecite{Dasso06}.
Then, for the out-bound branch we compute $F_z^{out-bound}$ as
(idem for $F_z^{in-bound}$, but with the integration between
$t_{in}$ and $t_{\rm center}$):
  \begin{equation}
   F_z^{out-bound} = 2 \pi \int_{t_{\rm center}}^{t_{\rm out}}
         B_{z,cloud}(t') \ (x(t')-x_{\rm center}) \ V_{x,cloud}(t') \rmd t'
   \label{Fz_direct}
  \end{equation}
As in previously studied MCs (e.g. \citeauthor{Mandrini05}
\citeyear{Mandrini05}, \citeyear{Mandrini07}; \opencite{Attrill06})
$F_z$ is one order of
magnitude lower than $F_{\phi}$ (see Table~\ref{tab_flux}).

\subsection{Structure of the ICME}
        \label{direct_method_ICME_structure} 

   The analysis of $F_y/L_{\rm in}$ in
Section~\ref{direct_method_structure} indicates that this MC is
not formed only by a flux rope.  Some of the MC characteristics
(see Figure~\ref{fig_BV_gse}), such as a very unusually high magnetic
field with a low variance, a low $\beta_p$, and a strong
expansion, continue well after the rear boundary of the flux rope
(``out$_1$'', ``out$_2$'', and ``out$_3$'').
We discuss below the most plausible
physical scenario to create such magnetic structure.

  There is an extended region where $B_{y, \rm cloud}$ has still a
negative and coherent behavior from position ``out$_3$'' to ``back'',
so the accumulated flux keeps increasing monotonously
(Figure~\ref{fig_Fy}).  This behavior was also found in the 18~Oct.
1995 MC and it was interpreted as the trace of an original larger magnetic
flux rope whose front was partially reconnected with the overtaken
magnetic flux, as shown in Figure~6 of \inlinecite{Dasso06}.  In the
example analyzed here, this interpretation has even more support from
the data, since part of the overtaken flux is still present in front of the MC
and we can estimate when reconnection started.

   There is a coherent negative $B_{y, \rm cloud}$ field just in front of
the MC (from $\approx$~18:00~UT to 20:30~UT, Figure~\ref{fig_BV_cloud}).
We interpret this as the remnant of the magnetic flux which reconnected
with the original flux rope. At the time of observations, the MC
overtakes this structure with a velocity difference of $\approx
40$~km/s; so reconnection is driven. Such velocity difference
is crucial for reconnection efficiency since its rate increases
with a larger velocity difference \cite{Schmidt03}. Furthermore,
\inlinecite{Burlaga95} and \inlinecite{Farrugia01} have found that
magnetic holes, such as the one preceding
the MC analyzed here (top panel of Figure~\ref{fig_BV_gse}),
are associated with magnetic reconnection.

   Taking into account the previous
reconnection scenario, the back region, where
$B_{y,\rm cloud}$ has still a coherent negative value, is simply the
magnetic flux at the periphery of the original flux rope. This back region
keeps the properties of typical MCs without fitting in the standard flux rope
models.  If the front reconnected, this back region is connected to the
solar wind field; so, the magnetic field direction can change
because of the propagation of Alfven waves, as observed here with the reversal
of $B_{z,\rm cloud}$ (Figure~\ref{fig_BV_cloud}). However,
if there is no other reconnection process,
the flux of $B_{y, \rm cloud}$ cannot be
removed from the back of the MC; in fact, we find no other field
to allow such reconnection. Thus, the closed (flux balanced)
flux rope observed at 1 AU is embedded in a larger structure
which includes an extended back region. This flux rope was part
of a larger one that was partially pealed at its front because
it reconnected with its environment.

\subsection{Clues for magnetic reconnection}
        \label{direct_method_reconnection} 

  The high velocity of the MC ($V_c \approx$~800~km/s), compared
to its surroundings ($\approx$~600~km/s) implies a progressive
extension of the back region (between boundaries ``out$_3$'' and
``back'' in Figure~\ref{fig_BV_cloud}).  Assuming that the relative velocity,
$V_c-V_{\rm back}$, was similar at earlier times and that the
$B_{y, \rm cloud}$ component (in the back region) had
initially a value similar to that in the
rear of the MC (because of pressure balance), we
can estimate the period of time $\delta t$ between the start of
reconnection and the MC observations.
The range of time between the observation of the two extremes of
the expanding region (between ``out$_3$'' and ``back'')
is $\tau_{\rm expansion} \approx$~18~h.
This corresponds to a spatial extension
of $\sim \tau_{\rm expansion} V_{\rm back}$
when the region was observed.
So, the back region expands in size by $\tau_{\rm expansion} V_{\rm back}$
since the start of reconnection at the MC front.
This extension has its origin in the relative velocity between the MC
and the back region (which becomes lower because of its
magnetic connection to the solar wind). From the beginning of
the reconnection at the MC front, starting a lapse of time $\delta t$
earlier than the observations, the ``back'' boundary progressively
separates from the MC with a relative velocity $V_{\rm MC}-V_{\rm back}$.
Assuming that this relative velocity was not changing drastically during
the MC transit, the back region expanded by
$\approx (V_{\rm MC}-V_{\rm back}) \delta t$. Equating the
previous estimations of the back region extension, we find:
    \begin{equation} \label{t-reconnection}
    \delta t \approx \frac{ \tau_{\rm expansion} V_{\rm back} }
                       { V_c-V_{\rm back} }
             \approx 18 \times 600 / 200 \approx 54~{\rm h} \,.
    \end{equation}

  The solar event that is the most probable source for the MC
studied here occurred in AR 10696. It is a multiple event that
starts with a steep rise in GOES light curve
at $\sim$ 15:50 UT on Nov. 7, and has two clear peaks, one at $\sim$ 16:00 UT and
the other at $\sim$ 16:35 UT. At the time of the second peak a
large two-ribbon flare was observed within AR 10696, but also
two H$\alpha$ ribbons were seen at both sides of an erupting
trans-equatorial filament extending from AR 10696 to AR 10695 at the
southwest (see \opencite{Harra06}). During this intense event that
reached class X2.0 in soft X-rays the full neutral line, which
formed a switch-back, erupted.
The CME observed in the Large Angle and Spectroscopic Coronagraph
(SOHO/LASCO) C2 on the Nov. 7 at 16:54 UT
is associated with this multiple flare involving AR 10696 and the
erupting trans-equatorial filament to the southwest.
\inlinecite{Longcope07} have proposed that
the source of the MC is the AR eruption,
while \inlinecite{Harra06} have considered both
possibilities either the AR or the trans-equatorial filament eruption.
We discuss below these scenarios in view of our IP analysis.

  Taking $\sim$~15:50~UT on Nov. 7,
as the start time of the solar event, and an arrival time for
the MC front at $\sim$ 20:30~UT on Nov.~9, we obtain a transit time
of $\sim$~52~h, comparable to the transit time ($\sim$~47~h) computed assuming
a constant MC velocity of -800~km/s and the spacecraft
located at L1. Both estimations
are comparable to $\delta t$ (Equation
\ref{t-reconnection}); this implies that reconnection started when the
flux rope was close to the Sun.

 What could be the magnetic structure present in front of the flux
rope, probably already from its origin in the corona? The
field in front of the MC is oriented mostly southward with an
average field intensity of $\sim$ -30 nT, though there is
a structure of northward oriented field with a temporal
length between 19:54 UT and 20:01 UT on Nov. 9,
an average field strength
of $\sim$ -15 nT, and a spatial extension of $\sim$ 0.002 AU
(taking an average velocity of
$\sim$ -800 km/s). Analyzing the MDI magnetogram on Nov.~7, the most probable
origin of the southward oriented structure is
the large scale, nearly potential field of
AR 10696. If the source of the MC is the AR eruption; then, the
core of the AR should become kink unstable (as
it was observed in an eruption which occurred
in the same AR three days later \cite{Williams05}).
In the kink instability, part of the twist is
transformed into writhe implying a strong rotation of the flux rope.
Taking into account
the MC orientation found in Section~\ref{direct_method_orientation},
this rotation should be $\sim$ 160$^\circ$.  The MHD simulation
of \inlinecite{Gibson04} gives a writhing of the flux tube of $\sim$ 120$^\circ$
(before it reconnects with the overlying field).
Numerical simulations by \inlinecite{Torok05} confirm this using a different
approach and, indeed, the writhing could be as large as a
rotation of 160$^\circ$ depending on the properties of the
overlying field (T{\"o}r{\"o}k, private communication). So the amount of
rotation and its direction (for a left-handed flux rope), if the
MC flux rope comes from the AR, is coherent
with recent MHD simulations of kink unstable flux ropes.

Interplanetary scintillation observations suggest that
the material from the core of the AR was ejected
primarily northward and, thus, it could remain
unobserved at 1 AU (see \opencite{Harra06}).
In this scenario, and if the source of the MC is the trans-equatorial
filament eruption,
the direction of the trans-equatorial
filament and the cloud axis differ by 67$^{0}$ (see Figure 12
in \opencite{Harra06}). Moreover, the sense of
rotation from the filament to the cloud axis is opposite to the
one expected from the development of a
kink instability in a region of left-handed magnetic helicity.
As discussed by \inlinecite{Harra06}, it may happen that the kink instability
has not played a role far away from the strong fields of the AR.
The northward oriented structure observed
between 19:54 UT and 20:01 UT on Nov. 9 could be related to the northward
oriented trans-equatorial loops observed above the filament.
However, in both scenarios reconnection should be forced between the
ejected flux
rope and the above AR arcade field from the beginning of the eruption.
This explains the origin of a nearly antiparallel field (neglecting the
short period of northward directed field) in front of the flux
rope from the beginning of the launch, as indicated by
a value of $\delta t$ similar to the transit time.

  Our previous discussion has implications for the reconnection rate in a
collisionless plasma.  While the erupting flux rope pushed against an
overlying nearly anti-parallel magnetic field, we still
observe part of this overlying flux in front of the MC at 1~AU!  Of
course we have no way to determine how strong was the forcing, nor the
time dependence of the reconnection rate during the flux rope transit.
Still this observation is a clue that magnetic reconnection is not
efficient in a collisionless plasma, as expected from classical theory.
Some recent observations show direct evidence of magnetic reconnection
in a collisionless plasma as the solar wind \cite{Gosling05}.
Some numerical simulations show
that the Hall effect can increase the reconnection rate above the
classical rate (e.g., \opencite{Morales04}). Our observations set a
constrain on the reconnection rate, which can be quantified only
after a numerical modeling of the flux rope ejection and
transit to 1~AU.

   From the computed accumulated flux (Figure~\ref{fig_Fy}) we can
estimate the relative amount of reconnected flux from the original
flux rope ejected from the Sun as $\approx 1.25/7.4 \approx 17$\%,
assuming $L(t)=L_{in}$ (left panel).
With the expansion length given by Equation~(\ref{Length_evol}), the
flux present in the back part, which is the fraction of reconnected flux,
is $\approx 30$\% (however it is not obvious that we can
use Equation~(\ref{Length_evol}) for the back part where the
field is connected to the solar wind).
The relative amount of flux present in front of the MC is $\approx 0.6/7.4
\approx 8$\%.  This implies that the relative amount of stable
(not kinked) magnetic
flux, arch-like, above the erupting flux rope is in the interval $[25,38]$\%
of the azimuthal flux in the MC.

   Finally, a fraction of the $B_{y, \rm cloud}$ magnetic flux
observed after the boundary ``in''
is expected to be reconnected when the spacecraft exits the flux rope.
To estimate an order of magnitude for this flux, let us suppose that the
reconnection rate was comparable to its mean value during the transit
from the Sun. The amount of reconnected flux during the MC observing time
is the ratio of the crossing time ($\approx 14$~h) over the transit time from
the Sun ($\approx 52$~h) times the flux reconnected from the Sun.
Then, $\sim$ 5\% of the original flux was reconnected during the
MC crossing time. So, the fraction of flux reconnected between $t=t_{\rm in}$
and $t_{\rm out}$ is expected to be an amount much lower than the uncertainties
in the flux estimations (see Table~\ref{tab_flux}).

\section{Results using fitted models}
\label{Sect_Models}

This cloud presents a large velocity difference between its
front and rear parts (fourth panel in Figure~\ref{fig_BV_gse}).
This velocity difference is the consequence of a large MC size
and a significant expansion.
Both conditions imply that the
magnetic field is observed at significantly different times
during the MC crossing, so at times when the MC
has significant different sizes.
As a consequence of magnetic
flux conservation, this implies the observed
decay of $B$ with time (upper panel of Figure~\ref{fig_BV_gse}).
Then, $B_{z,cloud}$, and even more $B_{y,cloud}$
(since it is stronger close to the flux rope borders)
presents a remarkable asymmetry between the in- and out-bound
branches.
Moreover, the center of the cloud is observed before the
central observing time for the full structure, as expected
for an spatially symmetric expanding object.

In this section we compare the observations to fitted models
describing the evolution of the magnetic field assuming an isotropic
self-similar expansion. The observed velocity is used to derive
the expansion rate, which is then used in the expansion magnetic model.
We also quantify the magnetic fluxes from the fitted models,
and compare them with the results obtained from a classical static model
and the direct method.

\subsection{Expansion Model}
\label{expansion_models}

   We assume an isotropic self-similar expansion of the MC, where
all distances are multiplied by a factor $f(t)$:
 \begin{equation}
  \vec{r}(t) = \vec{r}_0 f(t)   \,,
    \label{r_self_similar}
 \end{equation}
where $f(t_0)=1$ and $\vec{r}_0$ is the position
of a given element of fluid at some reference time $t_0$.
Since the MC front boundary, called ``in'', is well defined,
we select this as the reference time, $t_0=t_{\rm in}$.
Each element of fluid is labeled by its
$\vec{r}_0$ value, then the $\vec{r}_0$ coordinate is a marker of
each element of fluid (Lagrangian coordinates).

We assume below that the flux rope size increases
linearly with time. Then, $f(t)$ can be written as:
 \begin{equation}
   f(t) = 1+(t-t_{\rm in})/T \,,
  \label{f_linear}
 \end{equation}
where $T$ is the time of expansion. If Equation~(\ref{f_linear})
would be valid for all the transit time from the Sun, then
the flux rope size would vanish at time $t=t_{\rm in}-T$, so $T$ would also be
approximately the transit time
from the Sun to the position observed at $t_{\rm in}$. In practice, we use
Equation~(\ref{f_linear}) only during the MC crossing; so, we
are only assuming a local linear increase of the size with time.
Depending on the evolution during the transit from the Sun,
$T$ could be different from the transit time.

   The time evolution for the plasma velocity $\vec{V}$ in the cloud frame
comes from Equation~(\ref{r_self_similar}) doing the temporal derivative
and keeping $\vec{r}_0$ fixed (so following a plasma element):
 \begin{equation}
  \vec{V}(t) = \frac{\vec{r}}{T+t-t_{\rm in}} = \frac{\vec{r}_0}{T} \,.
    \label{V_r}
 \end{equation}
The first equality gives the velocity at a given time $t$,
so it is proportional to the distance. The velocity decays with time,
but this decay is only apparent since when one follows a given
plasma element, defined by $\vec{r}_0$, the velocity is in fact
constant (second equality). Then,
Equation~(\ref{f_linear}) implies that there are no forces acting on
any plasma element (free expansion). This is an ``a posteriori''
justification of Equation~(\ref{r_self_similar});
for an isotropic expansion, an initial force-free
field stays force-free.

The observed speed is the sum of the expansion speed and of the global
speed of the MC. During the time of observation, we can assume
that the MC is globally moving at a constant speed $-V_c$ (velocity of
the MC center) along $\hat{x}_{\rm cloud}$
because the aerodynamic drag significantly affects
$V_c$ only in a time scale comparable or larger than
the transit time to 1~AU. We also assume that the spacecraft
is at rest. Then, the velocity component along $\hat{x}_{\rm cloud}$ is:
 \begin{equation}
  V_{x} = -V_c + V_c \ \frac{t-t_c }{T+t-t_{\rm in}} \,.
    \label{V_x}
 \end{equation}
This equation is fitted to the data to derive $V_c$, $t_c$
(the time when the spacecraft crosses the cloud center),
and $T$.
It is noteworthy that the nearly linear observed $V_{x}$ is a
consequence of the constant speed $V_c$, and not of the
linear expansion with time assumed in Equation~(\ref{f_linear}).
The assumed expansion introduces only a non linear correction
in $(t-t_{\rm in})$ because the crossing time, $t_{\rm out}-t_{\rm in}$,
of the MC is small compared to the time of expansion, $T$ (a further
correction would come from the non-linear development of a general
$f(t)$). This justifies the use of Equation (\ref{f_linear})

\begin{figure}
 \centerline{\includegraphics[width=0.8\linewidth]{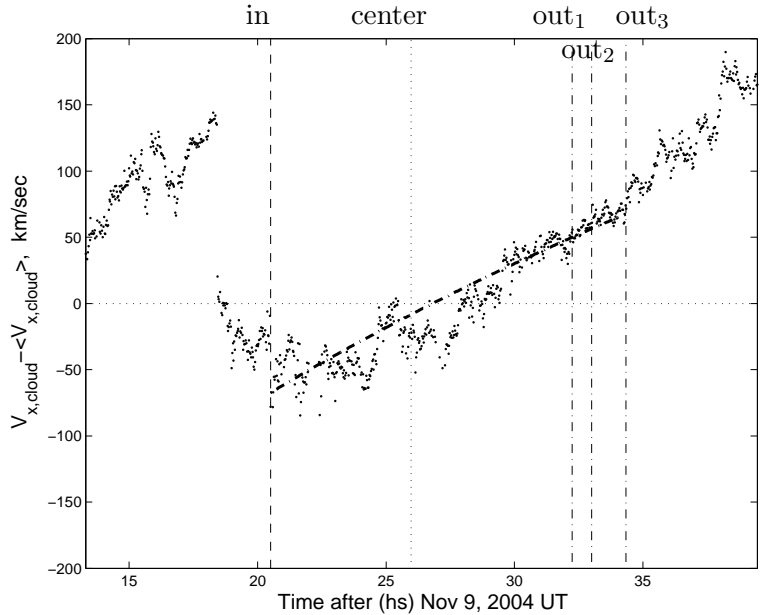}}
 \vspace{-0.66\linewidth}
 \centerline{\hspace{0.345\linewidth} in
             \hspace{0.065\linewidth} center
             \hspace{0.09\linewidth} out$_1$
             \hspace{0.01\linewidth} out$_3$  \hfill
            }
 \centerline{\vspace{0.01\linewidth} \hspace{0.42\linewidth} out$_2$
            }
  \vspace{0.62\linewidth}
\caption{
Observed (dots) and fitted (thick dash-dotted line) radial
velocity profile ($V_{x,cloud}-<V_{x,cloud}>$). Vertical
lines mark the same times as in Figure~\ref{fig_BV_gse}.}
\label{fig_Vr_model}
\end{figure}

\subsection{Results for the expansion}
\label{results_expansion}

The observed $V_{x,cloud}$ has a globally linear variation with time
within the MC (Figure~\ref{fig_Vr_model}). To better show
its variation, we subtract the mean value of the
velocity computed inside the range of positions ``in'' to ``out$_3$'', \linebreak
$<V_{x,cloud}>=-794$~km/s.
When the impact parameter ($p/R$) is small,
as it is in this MC, this speed represents
the radial velocity with respect to the cloud axis, such that
for a cloud in expansion, $V_{x,cloud}-<V_{x,cloud}>$ is negative
before the spacecraft reaches the cloud axis and positive after that.

From a least square fitting of Equation~(\ref{V_x})
to the velocity data
inside the ``in'' to ``out$_3$'' boundaries, we obtain
$T \approx T_3 = 73$~h ($\approx 3$~days). We find a slightly
longer time of expansion $T \approx T_1 = 79$~h ($\approx 3.3$ days)
when the fit is restricted to the interval between the
``in'' and ``out$_1$'' boundaries, and we get
$T_2 = 77$~h ($\approx 3.2$ days) for boundaries ``in'' to ``out$_2$''.
Considering the self-similar expansion given
by Equation~(\ref{f_linear}), this implies that the MC has expanded
from the time $t_{\rm in}$ by a factor $1.19$ and $1.15$
when the spacecraft crossed ``out$_3$'' and ``out$_1$'',
respectively.

Both times of expansion, $T_3$ and $T_1$, are longer than the transit
time from the Sun
($\approx 52$~h, Section~\ref{direct_method_reconnection}).
Taking an approximately constant global velocity $V_c$, the distance
of the MC front to the Sun increases by the factor
$D(t)/D_{\rm in} \approx 1+ (t-t_{\rm in})V_c /D_{\rm in}$
(see Equation~\ref{Length_evol}). Doing the ratio of
the measured expansion rate variation ($\frac{df(t)}{dt}$)
to $\frac{dD(t)}{dt}$, we obtain an undimensional
factor $D_{\rm in}/(V_c  T)$. This factor is  0.64 and
0.59 for boundaries ``out$_3$'' and ``out$_1$'', respectively.
Considering the orientation
of the MC ($\theta=-10^\circ$ and $\varphi=275^\circ$), the velocity in
Figure~\ref{fig_Vr_model} is measured mainly across the flux rope. Then,
the MC is expanding at
a significantly smaller rate radially than what we expect assuming the
isotropic expansion in Equation~\ref{Length_evol}.

 Assuming that the spacecraft trajectory is close enough to the MC axis,
$p/R<<1$, the observed velocity lets us compute the radius
of the flux rope since the distance crossed ($\int V_{x,\rm cloud} \dt$)
is $R_{\rm in}+R_{\rm out}= R_{\rm in} (1+f(t_{\rm out}))$.
Taking the boundary ``out$_3$'', we find $R_{\rm in}=0.12$~AU
and $R_{\rm out3}=0.14$~AU; while for boundary ``out$_1$'' we find smaller
radii, $R_{\rm in}=0.10$~AU and $R_{\rm out1}=0.12$~AU.
If we do not consider the expansion, we get the mean values
because the distance crossed is simply $2 R$.

\subsection{Magnetic field models}
\label{magnetic_models}

  Lundquist's static model \cite{Lundquist50} is a classical
linear force free configuration ($\bar{\nabla} \times \bar{B}
= \alpha \bar{B}$, with $\alpha$ constant).
From the conservation of the magnetic flux during
the expansion, and assuming a Lundquist's field
at a given time, it is possible to derive an
expansion field model
(see, e.g., \opencite{Shimazu02}, \opencite{Berdichevsky03}).
We will also consider a more general model where the amplitude
of the azimuthal and axial components are independent.
This represents an approximation for a flux rope with an oblate
cross section (see \opencite{Vandas03} for an exact solution). We
consider this modified expansion Lundquist's model to keep
the same functional dependence of the field components, so the
difference between the three models is only the number of free
parameters (and the physics involved).

 The three models are described by the equations:
  \begin{eqnarray}
   B_r(r,t)      &=& 0
       \,, \label{B_r} \\
   B_{\phi}(r,t) &=& B_{\rm in,\phi} \ {\rm f}^{-2} \
                     J_1(\alpha_{\rm in} \ r/{\rm f})
       \,, \label{B_phi}  \\
   B_z(r,t)      &=& B_{\rm in,z}    \ {\rm f}^{-2} \
                     J_0(\alpha_{\rm in} \ r/{\rm f})
       \,, \label{B_z}
  \end{eqnarray}
where $B_{\rm in,\phi}$, $B_{\rm in,z}$, and $\alpha_{\rm in}$ are the field
and $\alpha$ values when the spacecraft enters the MC at $t=t_{\rm in}$.
For the static model ${\rm f}=1$ and
$B_{\rm in,\phi}= B_{\rm in,z} = B_{\rm in}$, so there are only two free
parameters,
$B_{\rm in}$ and $\alpha_{\rm in}$. For both expansion
models ${\rm f}=f(t)$ is function of time as given by
Equation~(\ref{f_linear}).
For the expansion Lundquist's model,
$B_{\rm in,\phi}= B_{\rm in,z} = B_{\rm in}$,
so there are also two free parameters as for the static model.
For the modified model, there is an extra free parameter; the parameters
are: $B_{\rm in,\phi}$, $B_{\rm in,z}$ and $\alpha_{\rm in}$.
Notice that for all models $B_z$ is not forced to vanish at
the MC boundaries (this lets $\alpha_{\rm in}$ as a free parameter).

  Because for a fixed time $t$, each of the components
$B_{\phi}$ and $B_z$ have a spatial dependence as in
Lundquist's model, the equations for the
magnetic fluxes $F_z$ and $F_\phi$ are the same as
in \inlinecite{Dasso06}, but now $R=R(t)=R_{\rm in}\ f(t) $ and
$L=L(t)=L_{\rm in}\ f(t)$ with $R_{\rm in}$ and $L_{\rm in}$ the radius
and the length of the
cylinder at time $t=t_{\rm in}$.
Then, $F_z$ and $F_{\phi}$ are, as expected, constants of motion,
because the increase in $L(t)$ and $R(t)$ cancels the decay of
the field components and of $\alpha$.
They both simply write:
  \begin{eqnarray}
   F_z &= & \frac{2 \pi B_{\rm in,\phi} R_{\rm in} J_1(\alpha_{\rm in} R_{\rm in})}{\alpha_{\rm in}}
       \label{Fz_RA} \,, \\
   F_{\phi} &= & \frac{B_{\rm in,z} \, (1-J_0(\alpha_{\rm in} R_{\rm in}))}{\alpha_{\rm in}} L_{\rm in}
       \label{Fphi_RA} \,.
  \end{eqnarray}

\begin{figure}
\centerline{\includegraphics[width=1.0\linewidth]{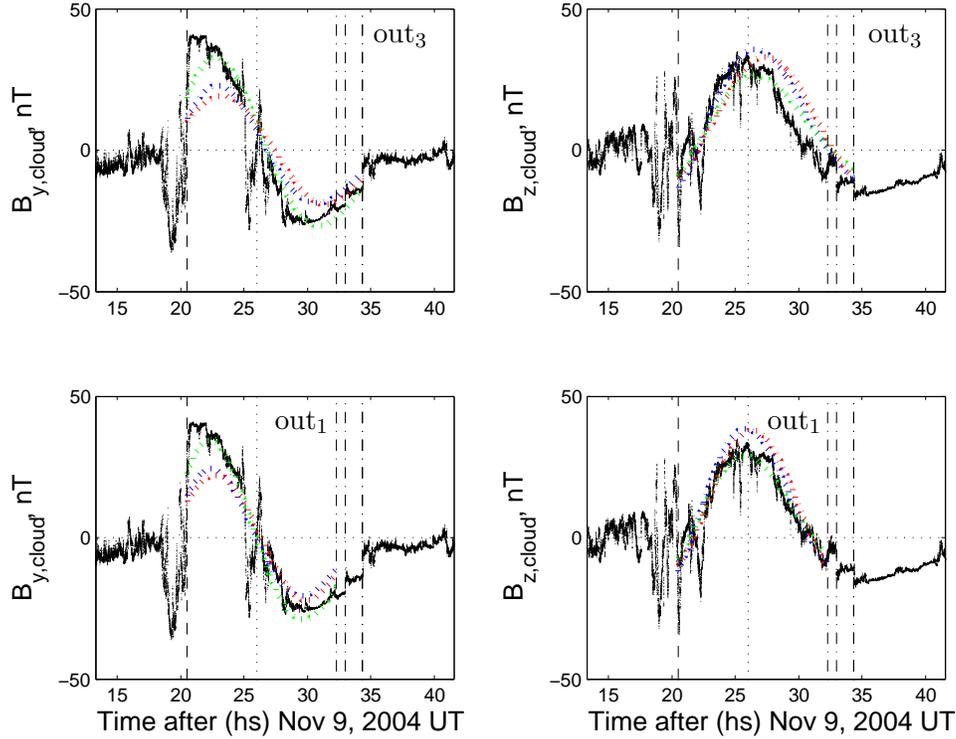}}
\vspace{-0.77\linewidth}
\hspace{0.36\linewidth} out$_3$
\hspace{0.40\linewidth} out$_3$
\vspace{0.37\linewidth}  \hfill
\centerline{
\hspace{0.14\linewidth} out$_1$
\hspace{0.44\linewidth} out$_1$
}
\vspace{0.36\linewidth}
\caption{$B_{y,cloud}$ (left panels) and $B_{z,cloud}$
(right panels). Observations (dots), cylindrical
dynamical model (red dotted lines),
cylindrical static model (blue dotted lines),
and modified model (see text, green dotted lines).
Upper panels show the fitting using the end time
as Nov. 10, 10:20 UT (out$_3$),
lower panels use Nov. 10, 8:15 UT (out$_1$) as the end time.
Vertical lines mark the same times as in Figure~\ref{fig_BV_gse}.}
\label{fig_B_fitting}
\end{figure}

\subsection{Results for the magnetic field}
\label{comparison_models}

   In this section we fit the MC observations with
the three models described in Section~\ref{magnetic_models}.
We fix the orientation
to the one given in Section~\ref{direct_method_orientation}
($\theta=-10^\circ$ and $\varphi=275^\circ$). This allows us
to test the effect
of including the expansion and the decoupling of azimuthal/axial field
from the problem of finding the MC axis.

  We first use the data in between  ``in'' and ``out$_3$''.
We use a nonlinear fitting routine to fit the models
presented in Section~\ref{magnetic_models},
assuming $p=0$, to the observations of $B_{y,cloud}$ (which corresponds
to $\pm B_{\phi,cloud}$) and $B_{z,cloud}$.
  The static model cannot reproduce the observed
asymmetry due to the decay of the field and the shift
of the position of the cloud center,
because of its intrinsic symmetry (top panels of Figure~\ref{fig_B_fitting}).
Both the static and expansion Lundquist's models overestimate
the axial field, $B_{z,cloud}$, near the cloud center and underestimate the
azimuthal field, $B_{y,cloud}$, near the boundaries.
When the extra freedom $B_{\rm in,z} \not= B_{\rm in,\phi}$ is considered, the
model can reproduce significantly better the observations.

A quantitative comparison between the different models
is given by $\sqrt{\chi^2}$, where $\chi^2$ is the time average
of $(\vec{B}_{model}-\vec{B}_{observations})^2$. We find
$\sqrt{\chi^2}$=14:16:19~nT, for the modified model, expansion and static
Lundquist's models, respectively.
For the three models a shift of the positions where $B_{y,cloud}=0$
and where $B_{z,cloud}$ is maximum
to a time later than in the observations is present
(top panels of Figure~\ref{fig_B_fitting}).

  We also explore the two earlier rear boundaries:
``out$_1$'' and ``out$_2$''.
A significant better fit is found with the boundary at
positions ``out$_1$'' and ``out$_2$'' (lower panels
in Figure~\ref{fig_B_fitting} show the fitting
for ``out$_1$'', very similar fitted curves
are obtained for ``out$_2$'').

A quantitative comparison between the fits is given in the form \linebreak
$\sqrt{\chi^2_{\rm out,1}}$:$\sqrt{\chi^2_{\rm out,2}}$:$\sqrt{\chi^2_{\rm out,3}}$
in units of~nT.
For the static model we obtain 16:17:19, for the expansion model
15:15:16, and for the modified model 12:12:14.

If the boundary is located at ``out$_2$'', the modeled cloud center
corresponds now to Nov. 10 at 02:16UT, 14 minutes later than
the center given by the observations and the direct method in
Section~\ref{direct_method_structure} (dotted line,
Nov 10, 02:02UT). For a rear boundary at ``out$_1$'',
the modeled cloud center corresponds to Nov. 10 at 01:58UT,
only 4 minutes earlier than the value obtained using the
direct method.

  Let us now analyze the differences between the observations and the
best model
(modified Lundquist's model with boundary ``out$_1$''). The model follows
globally
well the observations, except for the central reversal of $B_{y,cloud}$
and close to the boundaries. This central reversal cannot be taken into
account by the
model (see Section~\ref{direct_method_structure} for an analysis of
this feature).  Close to the boundaries,
the asymmetry of the model is not as large as the observed one; the
isotropic expansion model
gives $B_{y, \rm out_1}/B_{y,\rm in}= f^{-2}(t_{\rm out_1}) \approx 0.75$,
while the observations give
$\approx 0.5$. Due to the crossing geometry, the data
mainly reflect the radial expansion velocity.
Let us now consider a refinement of the above model, the radial
expansion is still given by $f(t)$, while the axial expansion is
rather given by $g(t)=L(t)/L_{\rm in}$ (Equation~(\ref{Length_evol})).
$B_{y,cloud}$ is affected both by the radial and the axial expansion.
The conservation of the azimuthal flux
gives: $B_{y}(t)=B_{y,\rm in}/[f(t)g(t)]$.
In the fitting of the models, we have assumed $f(t)=g(t)$, but in fact
$D_{\rm in}/V_{c} <T$ so that $g$ is slightly larger than $f$.
With $D_{\rm in}/V_{c} \approx 47$~h and $T\approx 79$~h, we find
$f_{\rm out_1}=1.15$ and $g_{\rm out_1}=1.25$, which gives
$B_{y, \rm out_1}/B_{y,\rm in} \approx 0.69$ (rather than $0.75$ with
an isotropic expansion), a value closer to the observed value
($\approx 0.5$) but still larger.

  A spatial asymmetry between the in- and out-bound branches, which cannot be
attributed to the expansion, is present in the observations
(Figure~\ref{fig_B_fitting}).
Indeed, the observed $B_{y,cloud}$ does not have the expected decrease
towards the front boundary (which is present in the model and also in the
observations towards ``out$_1$'', Figure~\ref{fig_B_fitting}).
At the MC front the magnetic field is expected to be compressed
(then, it is enhanced) by the dynamic
pressure of the overtaken plasma. The interaction with the surroundings
is likely to be at the origin of this extra asymmetry (on top of the expansion)
between the in- and out-bound branches.

\begin{centering}
\begin{table}[t]
      \begin{tabular}[h]{cccccrrrr}
        \hline
          & end: & \multicolumn{2}{c}{``out$_1$''} &
\multicolumn{2}{c}{``out$_2$''} & \multicolumn{2}{c}{``out$_3$''} \\
        Model or Method & &  $F_z$ & $F_{\phi}$ & $F_z$ & $F_{\phi}$ &
$F_z$ & $F_{\phi}$\\
        \hline
       static    && 7.4 & 60 & 7.7 & 60 & 7.4 & 60 \\
       expanding && 7.4 & 64 & 7.4 & 67 & 6.8 & 69 \\
       modified  && 6.4 & 91 & 6.8 & 96 & 6.2 & 100 \\
       direct without axial expansion    && 5.0 & 81 & 4.7  & 85 & 2.1 & 91
\\
       direct with axial expansion    && 5.0 & 90 & 4.7  & 95 & 2.1 & 102 \\
        \hline
\end{tabular}
\caption{
Magnetic fluxes present in the flux rope in units of $10^{20}$~Mx
using different fitted models or the direct method, for
three backwards boundaries (``out$_1$'', ``out$_2$'',
and ``out$_3$'' in
Figures~\ref{fig_BV_gse} and~\ref{fig_BV_cloud}), and for
orientation given by $\theta=-10^\circ$ and $\varphi=275^\circ$.
For the direct method we present an average between the values
obtained for the in-bound and out-bound branches
($F_z$ is not affected by including axial expansion).
The fluxes are not corrected for the flux lost
by magnetic reconnection during the MC travel from the Sun (since
only the direct method permits an estimation of the reconnected flux).
The three models are described by the same equations
but differ by the constraint set on the free parameters,
so the physics involved (Equations~\ref{B_r}-\ref{B_z}).
An initial length for the cloud $L_{\rm in}=1.5$~AU is
assumed.}
\label{tab_flux}
\end{table}
\end{centering}

From the fitted parameters and expressions given in
Equations~(\ref{Fz_RA})-~(\ref{Fphi_RA}), we obtain the values
for the magnetic fluxes using the different models (Table~\ref{tab_flux}).
We assume an initial length $L_{\rm in}=1.5$~AU as in
Section~\ref{direct_method_flux}.
From Table~\ref{tab_flux},
$F_z$ is estimated in the range [2.1-7.7]$\times 10^{20}$~Mx, and
$F_{\phi}$ in the range [60-102] $\times 10^{20}$~Mx.
Fixing the end boundary at Nov 10, 10:00 UT (only 20 minutes earlier
than boundary ``out$_3$''), and using the static classical
Lundquist's model, \inlinecite{Longcope07} reported $F_z = 7.2 \times 10^{20}$~Mx
and $F_{\phi}/L=41\times 10^{20}$~Mx/AU (thus $F_{\phi} = 62 \times 10^{20}$~Mx,
for a cloud length of $L=1.5$~AU, as assumed here).
These values, as expected, are very close to our present results with the static
model and boundary ``out$_3$''.
Our present results show that the expansion affects slightly the computed
fluxes (more $F_{\phi}$), while decoupling the fits of $B_{y, \rm cloud}$
and $B_{z,\rm cloud}$ has the largest effect.
For rear boundaries ``out$_1$'' and ``out$_2$'',
the estimations of both fluxes using
the modified Lundquist's model are
in close agreement with the results of the
direct method that consider an axial expansion,
in particular this agreement is much better
for $F_{\phi}$.
Considering the rear boundary ``out$_3$'', the value of
$F_{z}$ obtained from the direct method is lower than
the one from the modified Lundquist's model due to
the significantly negative value of $B_{z,cloud}$
beyond ``out$_2$'' (see Figure~\ref{fig_BV_cloud});
this contributes to decrease $F_{z}$ when integrating
the circular cloud section
between ``out$_2$'' to ``out$_3$''.
As discussed in Section~\ref{direct_method_orientation},
we believe that the rear boundary of the cloud should be
between ``out$_1$'' and ``out$_2$''.

\section{Summary and Conclusions}
\label{Conclusion}

  The ICME of Nov 9-10, 2004, was a complex event with a large expansion
and a strong magnetic field in the front decreasing monotonously
(almost linearly) with time. Earlier analysis considered that
two magnetic clouds (MCs) were located inside this ICME.
Latter studies concluded that only one MC was present.
We confirm this and precise
the orientation and boundaries of the flux rope using several methods.
We also find clues about the interaction of this MC with its surroundings.

  To facilitate the understanding of the physics involved in the MC it is
useful
to transform the data to the local MC frame where the axial and azimuthal
components of the magnetic field are decoupled. A classical method to determine
the MC axis is the minimum variance (MV) method, which takes into account
the different spatial behavior of the magnetic field components to find the
flux rope orientation. We minimize the effect of the strong expansion, which
implies a decreasing field magnitude with time, normalizing the field
at each data point.
This gives a range of possible orientations
(typically with a precision of the order of $\pm 20^\circ$).

  The determination of the orientation of the flux rope axis is improved using
the direct method. This method
is based on two main points: first, the flux rope is
topologically distinct from the surroundings, so it should generically
be bounded by a discontinuity of the magnetic field components
(presence of a current sheet),
and second, the same amount of azimuthal magnetic flux should
be present in the in- and out-bound branches of the cloud.
For the cloud of Nov. 9-10, 2004, the frontal discontinuity is well
defined, while three rear discontinuities are present,
called ``out$_1$'', ``out$_2$'', and ``out$_3$''.
They are separated by about 2~h compared to a MC duration
of about 14~h.
The azimuthal flux relates the frontal discontinuity to two
discontinuities at the rear using two different hypotheses
for the axial evolution: an expansion comparable to the radial
one gives a rear discontinuity at ``out$_2$'',
and a negligible axial expansion
gives a rear discontinuity at ``out$_3$''.
However, since the second case implies a reversal of
the axial field at the rear of the MC,
we conclude that the first discontinuity (``out$_2$'')
is associated to the frontal discontinuity.
This defines precisely the extension, as well as the orientation
angle $\theta $ (within $\pm 5^\circ$, $\theta=10\pm 5^\circ$)
of the flux rope.

  Fitting a model to the data is another approach to understand the
observed magnetic structure.
We have used three models that are based on the Lundquist's solution.
The first one is the classical static solution. The second one
includes a self-similar expansion with the same rate in the axial and radial
directions.
Finally, the third one also includes an isotropic expansion
and decouples the fit of the
azimuthal and axial field components to take into account the observed stronger
azimuthal component (a possible signature of a flat cross section).
The expansion rate is obtained fitting the model to the
observed plasma velocity. The best fits to the data are obtained when
the first and second discontinuities (labeled ``out$_1$'' and ``out$_2$'')
are used, in agreement with the results obtained with the direct
method.

  Comparing the results of the fitted models with the
direct method, using boundaries ``out$_1$'' and ``out$_2$'',
we find that the axial and azimuthal fluxes are in the ranges
[4.7-7.7]$\times 10^{20}$~Mx and [60-95]$\times 10^{20}$~Mx,
respectively.
The main limitation on the axial flux measurements is the
unknown shape of the cross section.
For the azimuthal flux, it is important to consider
the axial expansion.
Here the limitations are different, the shape of the cross section
is not important,
the main limitation is the distribution of the flux along the MC axis.
Finally, we confirm that the azimuthal flux is
one order of magnitude larger than the axial flux.

  After the large and coherent rotation of $\vec{B}$,
some typical MC characteristics are still present:
low level of fluctuations, strong expansion (observed
in the decay of $V$ and $B$), intensity of
magnetic field higher than the typical solar wind values,
low $\beta_p$. These are evidences that the MC
extends farther in the back of the flux rope. Part of this back
(from ``out$_3$'' to ``back'' in Figures~\ref{fig_BV_gse}
and~\ref{fig_BV_cloud}) shows a coherent
behavior of $B_{y,cloud}$, which we interpret as the signature of an
originally larger flux rope that was partially reconnected in its front
near the Sun, with the consequent flux removal.

  Reconnection of the cloud field with the overtaken solar wind field
is another source of underestimation of the original magnetic flux that
was launched from the Sun.
We have found that $\sim 17$~\% (with a $30$~\% as an upper bound)
of the azimuthal flux was lost in the front
of the MC during its travel from the Sun; this is much less than in the
Oct. 18-30, 1995 MC, where it was estimated to be about 57~\% \cite{Dasso06}.
Another difference between the MCs is that the Nov. 9-10 MC has
a back part moving at a speed significantly lower (by $\approx 200$~km/s)
than the flux rope (while the Oct. 18-20 MC was overtaken by a fast stream).
Then, the reconnected field progressively forms an extended
region in the back of the flux rope (with a weaker magnitude but still with
a smooth spatial variation).  From the extension and velocity difference
of this region with the flux rope, we have estimated that reconnection
started close to the Sun, possibly between the erupting
twisted flux tube (giving the flux rope) and the overlying arcade of
active region 10696.
Part of the arcade field is probably present
in front of the MC, with a nearly anti-parallel direction and a
significant velocity difference ($\approx 40$~km/s), indicating
that magnetic reconnection is not so efficient in the interplanetary space.

Reconnection in front of the MC
has several observational consequences, as follows:\\
  \indent First, it introduces an asymmetry in the observed magnetic
field. The remaining part of the flux rope is observed first followed by
an extended tail of weaker magnetic field (which is re-orientated
since it has changed its connectivity).
When this process is dominantly at work, this
implies the presence of a closed flux rope at the beginning of the ICME. \\
  \indent Second, depending on the solar launch direction, the spacecraft
could cross the flux rope or its laterally extended back part. In this
last case, one would detect some characteristics of the MC (such as a
coherent field and low $\beta$ values), but without the coherent rotation of the field.
Such crossing would be classified as an ICME (without MC). An example
of such observations with the two HELIOS spacecrafts is analyzed by
\inlinecite{Cane97} and another example using the ACE spacecraft
is analyzed by \inlinecite{Foullon07}.\\
  \indent Finally, the back flux connected to the solar wind
field makes the moving magnetic structure larger in the
transverse direction (orthogonal to the global motion). From its
mixed origin the back region is expected to move at a speed intermediate
between
the MC and the solar wind speed, as observed in the present analyzed MC.
Then, with a significant larger velocity than the surrounding medium, the back
region
is expected to have an effect on the frontal shock surrounding the ICME.
A larger transverse scale implies a forward shock at a larger distance in front
of the MC than the distance deduced by
its flux rope transverse size \cite{Farris94}.  Such large distance
has so far being interpreted as a flat flux rope \cite{Russell02}.
The consequence of reconnection in the front of the MC is an
alternative and/or complementary explanation which requires numerical
simulations to be quantified.

    The results obtained show the potentiality of combining several
methods of analysis, minimum variance, direct method and fit to the data.
This analysis will be done for other MCs to derive the variety
of possible physical scenarios and also to improve our
understanding of MCs and ICMEs.


\begin{acknowledgements}
The authors are grateful to L.K. Harra for organizing
the Sun-Earth Connection workshop at MSSL from which this contribution is
an outcome. This workshop was possible
thanks to a Phillip Leverhulm Prize.
S.D. would like to thank N.U. Crooker and
H. Elliott for helpful discussions.
This research has made use of NASA's Space Physics Data Facility (SPDF).
C.H.M. and P.D. acknowledge financial support from CNRS (France) and
CONICET (Argentina) through their cooperative science program (N$^o$ 20326).
This work was partially supported by the Argentinean grants:
UBACyT X329, PIP 6220 (CONICET), and
PICTs 03-14163, 03-12187 and 03-33370 (ANPCyT).
S.D. and C.H.M. are members of the Carrera del
Investigador Cient\'\i fico, CONICET.
M.S.N. is a fellow of CONICET.
\end{acknowledgements}

\end{article}
\end{document}